# Microwave Response of the Superconducting Diode Effect in Proximitized Bilayer Graphene Interferometers

Shili Yan[1], Rubén Seoane Souto[2,3], Yi Luo[1], Jeroen Danon [4], Haitian Su[1], Junze Zhang[1], Han Gao[1], Xingjun Wu[1], Ji-Yin Wang[1], H. Q. Xu[1,5*]

[1] Beijing Academy of Quantum Information Sciences, Beijing, 100193, China.

[2] Instituto de Ciencia de Materiales de Madrid (ICMM), Consejo Superior de Investigaciones Científicas (CSIC), Sor Juana Inés de la Cruz 3, Madrid, 28049, Spain.

[3] Quantum Advanced Research Center (QuARC), Consejo Superior de Investigaciones Científicas (CSIC), Sor Juana Inés de la Cruz 3, Madrid, 28049, Spain.

[4] Department of Physics, Norwegian University of Science and Technology, Trondheim, NO-7491, Norway.

[5] Beijing Key Laboratory of Quantum Devices, Peking University, Beijing, 100871, China.

***Corresponding author(s).** E-mail: hqxu@pku.edu.cn;

**Contributing authors:** yansl@baqis.ac.cn; ruben.seoane@csic.es; luoyi@baqis.ac.cn; jeroen.danon@ntnu.no; suht@baqis.ac.cn; zhangjz@baqis.ac.cn; gaohan@baqis.ac.cn; wuxj@baqis.ac.cn; wang_jy@baqis.ac.cn;

# Abstract

Microwave irradiation has emerged as a promising means to tune the superconducting diode effect (SDE) in Josephson junction devices. Previous experimental studies have mainly focused on the adiabatic-driving regime, in which the diode efficiency increases monotonically with microwave power and can approach the ideal value of unity. Beyond this regime, however, the microwave response of the SDE remains largely unexplored experimentally. In this work, we investigate the microwave response of the SDE in bilayer-graphene-based superconducting quantum interference devices (SQUIDs) under a broad range of driving frequencies. We show that increasing the driving frequency changes the response characteristics of the diode efficiency to microwave power—the dependence of the diode efficiency evolves from monotonic enhancement with increasing microwave power in the adiabatic regime to non-monotonic behavior beyond this regime, and ultimately to sign-reversal as well oscillatory characteristics at sufficiently high frequencies. We find that these experimentally observed frequency-dependent power response characteristics of the diode efficiency can be qualitatively captured by simulations based on the resistively shunted junction model using the device current-phase relations extracted from the experiments. These results establish SQUIDs made from bilayer graphene as a versatile platform for studying dynamic properties of superconducting junction devices.

**Keywords:** Superconducting diode effect, Bilayer graphene, Microwave irradiation

# 1 Introduction

Superconducting diodes, characterized by possessing the effect of nonreciprocal zero-resistance current transport, i.e., the superconducting diode effect (SDE), provide a set of superconducting analogues to classical diodes and thus new functional electronic components for low power-dissipative superconducting circuits [1, 2]. The SDE arises when certain symmetries in the superconducting devices are broken [2, 3]. A few strategies have been explored to realize SDE. Several approaches rely on intrinsic material properties to induce symmetry breaking, such as unconventional superconducting states [4–6], spin–orbit coupling [7–10], or valley polarization [11, 12]. Others exploit device-level symmetry engineering through, for example, superconducting quantum interference devices (SQUIDs) with higher harmonics in the current–phase relation (CPR) and asymmetry between the two arms [13]. These latter symmetry-engineering approaches are particularly attractive, as symmetry breaking at the device level allows for flexible material selections and a broad range of compatibility with diverse design requirements. The SDE in SQUIDs has been demonstrated in various material platforms in which clear tunability via both electrostatic gating and magnetic flux were achieved [14–20].

Microwave irradiation has been established as a powerful tool to probe and control superconducting effects [6, 8, 17, 18, 20–30]. In hybrid Josephson junction devices, microwave irradiation has recently been used to tune the SDE, with the microwave power being employed as an additional knob for controlling the diode efficiency [6, 8, 17, 18, 20, 23, 28, 29]. An example was demonstrated in Ge-based Josephson interferometers, where the microwave irradiation was shown to enhance the diode efficiency and drive it close to unity [17]. Similar microwave-induced enhancement of the diode efficiency was also observed in other hybrid superconducting systems [6, 28]. These observations can be understood within an adiabatic-driving picture. Here, "adiabatic" means that the microwave frequency is sufficiently low that, in determining the two edges of the zeroth Shapiro step, which correspond to the positive and negative switching currents under microwave irradiation, the junction response can be treated quasistatically at each instantaneous value of the microwave-modulated bias current. In this case, the magnitudes of the switching currents ($|I_{\mathrm{sw}}^{\pm}|$) are reduced approximately as $|I_{\mathrm{sw}}^{\pm}| - I_{\mathrm{ac}}$, where $I_{\mathrm{ac}}$ denotes the amplitude of the ac current through the device induced by microwave irradiation. Consequently, the diode efficiency increases monotonically with increasing $I_{ac}$ [23]. This simple relation also indicates that in the adiabatic regime, the microwave-power dependence of the diode efficiency is largely insensitive to the detailed form of the device CPR. This picture is, however, not expected to hold once the driving frequency is increased beyond the adiabatic regime. Theory predicts that, in appropriate parameter regions beyond the adiabatic limit, the diode efficiency could undergo polarity reversal and exhibit an oscillatory behavior as the microwave power is increased [18, 23]. However, experimental studies of the SDE under microwave frequencies beyond the adiabatic-driving regime remain largely unexplored [20].

In this work, we report on experimental studies of the microwave response of the SDE in SQUIDs made from bilayer graphene under a broad range of driving frequencies, covering both the adiabatic regime and that going beyond. The SQUIDs are fabricated using advanced van der Waals heterostructure assembly techniques [31, 32], with each consisting of two bilayer graphene (BLG)–aluminum (Al) Josephson junctions and a global back gate. Benefiting from the favorable transport properties of BLG and its excellent interfaces with superconducting electrodes [33–36], the devices exhibit a clear SDE that can be effectively tuned by both magnetic flux and electrostatic gate voltage. We have studied the microwave response of the diode efficiency in these devices at different driving frequencies. In the adiabatic regime, the diode efficiency increases monotonically with microwave power and approaches the ideal value of unity before reaching the first node of the zeroth Shapiro step, consistent with previous experimental observations in other hybrid systems [17, 28]. As the driving frequency is increased, the diode efficiency no longer exhibits the monotonic power dependence, indicating increasing importance of nonadiabatic phase dynamics. With further increasing driving frequency to sufficiently high values, the diode efficiency drops to zero, reverses its sign, and then shows an oscillatory behavior as the microwave power is increased. We analyze these observations and perform simulations based on the resistively shunted junction (RSJ) model using CPRs extracted from the experiments. We find that the experimentally observed microwave-power response of the diode efficiency of our bilayer graphene SQUIDs at different driving regimes can be qualitatively captured by the simulations.

# 2 Results

## 2.1 Basic Characteristics of the SDE in Bilayer Graphene SQUIDs

Two SQUIDs (SQUID-1 and SQUID-2) were fabricated and studied in this work. Each SQUID consists of two Josephson junctions (JJ1 and JJ2). The junctions were made from high-quality hBN/BLG/hBN van der Waals heterostructures and superconducting Ti/Al (4/30 nm) electrodes on a 300 nm-thick $SiO_2/Si^{++}$ substrate, where the heavily p-doped Si serves as a global back gate (see Methods for fabrication details). Figure 1a shows the optical image of SQUID-1 studied in the main article. In this SQUID, the junction JJ1 has a width of $W_1 = 4.3$ μm and a length (contact electrode separation) of $d_1 = 350$ nm, and the junction JJ2 has a width of $W_2 = 0.6$ μm and a length of $d_2 = 300$ nm. All basic characterization measurements were carried out using

DC techniques in a current bias, quasi-four-terminal circuit setup, as shown schematically in Fig. 1a, in a dilution refrigerator equipped with a uniaxial magnet at a base temperature of ~30 mK. Microwave irradiation was coupled to the device through an antenna positioned above the sample. The global back gate voltage ($V_{\mathrm{bg}}$) is set at 20 V, unless stated otherwise. Further measurement details are provided in the Methods section. All data shown in the main article are obtained from measurements of SQUID-1. Corresponding data from SQUID-2, which yield similar results, are presented in Supplementary Section 2.

We first present the basic characteristics of the SDE in SQUID-1. Figure 1b shows measured representative $V-I$ characteristics of the SQUID at nearly zero out-of-plane magnetic field $B\approx0$ (or magnetic flux $\Phi\approx0$). The solid and dashed traces in the figure correspond to upward and downward current sweeps, respectively. The two traces nearly overlap, indicating negligible hysteresis and that the device is in the overdamped regime without noticeable heating effects. $I_{sw}^{+}$ and $|I_{sw}^{-}|$ denote the switching currents extracted from the upward and downward current sweeps, respectively, and their difference reflects the presence of the SDE. In Fig. 1b, the SQUID exhibits a nearly identical switching current of $|I_{sw}^{\pm}|\sim1.6$ $\mu$A for the two current sweep directions, indicating a negligible SDE. Figure 1c shows the $V-I$ characteristics of SQUID-1 at a finite out-of-plane magnetic field of $B\approx6$ μT (or $\Phi\approx0.63\,\Phi_0$, where $\Phi_0=h/2e$ is the flux quantum). In contrast to Fig. 1b, a clear difference between $I_{\mathrm{sw}}^{+}$ and $|I_{\mathrm{sw}}^{-}|$ emerges, indicating the presence of the SDE with the diode efficiency, defined as $\eta=(I_{\mathrm{sw}}^{+}\text{-}|I_{\mathrm{sw}}^{-}|)/\,(I_{\mathrm{sw}}^{+}\text{+}|I_{\mathrm{sw}}^{-}|)$, reaching ~20 %. Figure 1d shows the differential resistance $dV/dI_{\mathrm{b}}$ of the SQUID as a function of the normalized external magnetic flux $\Phi/\Phi_0$ and bias current $I_{\mathrm{b}}$. Since no pronounced hysteresis is observed in the $V-I$ characteristics, only the data from one current-sweep direction are shown from now on unless otherwise specified. A clear periodic modulation of the switching currents is observed with a flux period of approximately one flux quantum $\Phi_0$, consistent with the expected SQUID interference pattern. From the interference period $\Delta B$, we extract an effective loop area $S=\frac{\Phi_0}{\Delta B}\approx227\ \mu\mathrm{m}^2$. This value is slightly larger than the geometric loop area ($\sim205\ \mu\mathrm{m}^2$) of the device, which could naturally be attributed to flux expelling and focusing and/or London penetration. The white arrows in Fig. 1d indicate the minima of $I_{\mathrm{sw}}^{+}$ and $|I_{\mathrm{sw}}^{-}|$. A clear offset in $\Phi/\Phi_0$ between these two extrema indicates the presence of the SDE. Figure 1e summarizes the extracted values of $I_{\mathrm{sw}}^{+}$ and $|I_{\mathrm{sw}}^{-}|$ as a function of $\Phi/\Phi_0$, showing that the difference between $I_{\mathrm{sw}}^{+}$ and $|I_{\mathrm{sw}}^{-}|$ is generally nonzero, except near integer and half-integer multiples of $\Phi_0$. Figure 1f displays the corresponding diode efficiency $\eta$ as a function of $\Phi/\Phi_0$. The efficiency exhibits flux-periodic oscillations with a period of one flux quantum $\Phi_0$, reaching its maximum negative and positive values at $\Phi_{-} \sim 0.32\,\Phi_0$ and $\Phi_{+} \sim 0.63\,\Phi_0$, respectively. This sign-reversal of $\eta$ reflects the phase-controlled nature of the SDE, arising from the interference between the two asymmetric Josephson junctions in the SQUID loop [13–17, 20]. The SDE can also be tuned by the global back gate, as shown in Supplementary Section 1, Fig. S1.

To further visualize the rectification behavior of the device, we apply a square-wave current to SQUID-1 with an amplitude $|I_\mathrm{b}| = A$ and monitor the output voltage at $\Phi \sim \Phi_{+}$, where $I_\mathrm{sw}^{+} > |I_\mathrm{sw}^{-}|$. When $|I_\mathrm{sw}^{-}| < A < I_\mathrm{sw}^{+}$, a finite dc voltage appears only when the current flows in the negative direction ($I_\mathrm{b} = -A$), while the system remains dissipationless for $I_\mathrm{b} = +A$. Figure 1g shows a representative segment of the measured voltage trace used for the histogram analysis in Fig. 1h. As seen in Fig. 1g, the measured voltage (red trace in the lower panel) alternates stably between two values, synchronized with the polarity of the applied current (blue trace in the upper panel). Figure 1h shows a histogram of the measured voltages constructed from 20,000 consecutively recorded data points over a total duration of approximately 6.6 hours. Here, two well-separated peaks in the voltage histogram, corresponding to the dissipationless ($V \approx 0$) and dissipative ($V \neq 0$) states, are clearly resolved, confirming stable and reproducible rectification behavior over time. All these results demonstrate the basic SDE functionality of the BLG–Ti/Al SQUID architecture.

### 2.2 Microwave Response of the SDE in Bilayer Graphene SQUIDs

We now investigate the response of the SDE in our bilayer graphene SQUIDs to microwave irradiation. In superconducting junction devices, such as Josephson junctions and SQUIDs, the microwave response is typically characterized by the appearance of Shapiro steps, well-known signatures of phase locking between the device and the external driving microwave. These Shapiro steps occur at voltages $V = \frac{Nhf}{2e}$, where $f$ is the microwave frequency. $N$ is integer for a device with a sinusoidal CPR. When higher-order harmonics are present in the device CPR, fractional Shapiro steps can appear, corresponding to non-integer values of $N$. Under microwave irradiation, the SDE manifests as a finite difference between the current magnitudes at the positive and negative edges of the zeroth-order Shapiro step.

We begin by presenting the microwave response of the SDE in SQUID-1 at a frequency of $f \sim 1.2$ GHz. Figures 2a–2d display the differential resistance $dV/dI_\mathrm{b}$ of the device as a function of the dc bias current $I_\mathrm{b}$ and microwave power $P$, measured at four different values of external magnetic flux, $\Phi \sim 0$, $0.5\,\Phi_0$, $\Phi_{-}$, and $\Phi_{+}$. The white dashed lines are guides to the eyes at $I_\mathrm{b} = 0$. In each color map panel, Shapiro steps appear as regions of minima in $dV/dI_\mathrm{b}$, with the central dark region corresponding to the zeroth-order step. For $\Phi \sim 0$ and $0.5\,\Phi_0$, the two edges of the zeroth-order step are symmetric with respect to $I_{\mathrm{b}} = 0$, indicating the absence of the SDE. In contrast, for $\Phi \sim \Phi_-$ and $\Phi_+$, the two edges become asymmetric with respect to $I_{\mathrm{b}} = 0$, signaling the SDE. Moreover, the SDE exhibits a clear dependence on the microwave power $P$. When one of the edges crosses $I_{\mathrm{b}} = 0$, the diode efficiency reaches 100%. For both $\Phi_-$, and $\Phi_+$, these crossings occur before the first node of the zeroth Shapiro step, consistent with the predicted behavior for a junction device in the adiabatic driving regime by the RSJ model [23, 28].

Next, we extract the diode efficiency $\eta$ as a function of microwave power $P$ from data in Figs. 2c and 2d. As shown in Fig. 2e, the absolute value of $\eta$ increases monotonically with $P$, approaching unity at high power. This behavior can be understood based on the RSJ model,

$$I_{\mathrm{dc}} + I_{\mathrm{ac}} \sin(\omega t) = I(\varphi) + \frac{\hbar}{2eR_n}\frac{d\varphi}{dt}, \tag{1}$$

where $\omega$ is the microwave angular frequency, $R_n$ is the resistance of the dissipative transport channel, $I(\varphi)$ is the total CPR of the SQUID, $I_{\mathrm{dc}}$ is the applied dc current drive, and $I_{\mathrm{ac}}$ is the amplitude of the ac current through the device induced by microwave irradiation. Now, the positive and negative current values at the two edges of the zeroth Shapiro step are denoted by $I_{\mathrm{sw}}^{+}(I_{\mathrm{ac}})$ and $I_{\mathrm{sw}}^{-}(I_{\mathrm{ac}})$. The diode efficiency is then given by $\eta(I_{\mathrm{ac}}) = \frac{I_{\mathrm{sw}}^{+}(I_{\mathrm{ac}})-|I_{\mathrm{sw}}^{-}(I_{\mathrm{ac}})|}{I_{\mathrm{sw}}^{+}(I_{\mathrm{ac}})+|I_{\mathrm{sw}}^{-}(I_{\mathrm{ac}})|}$. At sufficiently low microwave frequencies, the zeroth Shapiro step can be described quasistatically by considering the instantaneous bias current, which varies during each microwave cycle as $I_{\mathrm{dc}} + I_{\mathrm{ac}}\sin(\omega t)$. Assuming that heating and nonequilibrium effects are negligible, the zero-voltage state persists as long as the instantaneous bias current remains between $I_{\mathrm{sw}}^{-}(0)$ and $I_{\mathrm{sw}}^{+}(0)$, the negative and positive switching thresholds in the absence of microwave irradiation, respectively. In the equivalent washboard-potential picture, this condition corresponds that the phase remains trapped near an instantaneous local minimum of the slowly varying potential throughout the cycle. Once the instantaneous bias current goes beyond either threshold, the corresponding local minimum temporarily disappears, and the slowly varying drive affords the phase sufficient time to escape and complete a phase slip. The magnitudes of the positive and negative switching currents can therefore be expressed as $|I_{\mathrm{sw}}^{\pm}(I_{\mathrm{ac}})| = |I_{\mathrm{sw}}^{\pm}(0)| - I_{\mathrm{ac}}$, without explicitly solving the full time-dependent phase dynamics described by Eq. (1) [23]. As a result, the diode efficiency can be written as $\eta(I_{\mathrm{ac}}) = \frac{I_{\mathrm{sw}}^{+}(I_{\mathrm{ac}})-|I_{\mathrm{sw}}^{-}(I_{\mathrm{ac}})|}{I_{\mathrm{sw}}^{+}(I_{\mathrm{ac}})+|I_{\mathrm{sw}}^{-}(I_{\mathrm{ac}})|} \sim \frac{I_{\mathrm{sw}}^{+}(0)-|I_{\mathrm{sw}}^{-}(0)|}{I_{\mathrm{sw}}^{+}(0)+|I_{\mathrm{sw}}^{-}(0)|-2I_{\mathrm{ac}}}$. We assume that the ac current amplitude induced in the device is proportional to the square root of the applied microwave power. Since the microwave power $P$ is expressed in dBm, this assumption gives $I_{\mathrm{ac}} \propto 10^{P/20}$. Therefore, the diode efficiency can be recast as $\eta(I_{\mathrm{ac}}) = \eta(0)(1-\alpha 10^{\frac{P}{20}})^{-1}$, where $\eta(0)$ is the diode efficiency without microwave irradiation, and $\alpha$ is a proportionality constant related to the applied microwave power and the ac current generated in the device [28]. The solid curves in Fig. 2e show the fits to this expression by taking $\alpha \approx 0.93$, which is in good agreement with the experimental data. The fitted values of $\eta(0)$ are approximately $-16\,\%$ and $15\,\%$ for $\Phi \sim \Phi_-$ and $\Phi \sim \Phi_+$, respectively. Their magnitudes are slightly smaller than the corresponding values of approximately $18\,\%$ measured prior to microwave irradiation. This discrepancy could be attributed to an elevated electron temperature during the microwave measurements [23, 37]. Similar tuning behaviors are observed in SQUID-2, as shown in Supplementary Section 2.

We then present the microwave response of the SDE at a higher frequency. Figures 3a-3c display the measurement results at $f \sim 7.0$ GHz. Figures 3a and 3b show the Shapiro maps measured at $\Phi \sim \Phi_-$ and $\Phi \sim \Phi_+$, respectively. Here, well-defined integer and half-integer Shapiro steps are observed (see a few labelled low-order step regions in the figures). The appearance of the half-integer Shapiro steps indicates the presence of higher-order harmonics in the CPR of the device and is consistent with the fact that fractional Shapiro steps are more likely to appear at higher frequencies [38]. Unlike the low frequency (1.2 GHz) behavior shown in Fig. 2, the switching currents do not cross $I_\mathrm{b} = 0$ at $f \sim 7.0$ GHz, indicating that the diode efficiency does not reach unity at this higher frequency. To better illustrate how the diode efficiency $\eta$ evolves with $P$, we extract the switching currents $I_\mathrm{sw}^+$ and $|I_\mathrm{sw}^-|$ from Fig. 3a and 3b, and calculate the corresponding diode efficiencies $\eta$ as a function of $P$. The results are shown in Fig. 3c. The diode efficiency shows small initial increase with increasing $P$, followed by decrease to a broad dip at intermediate powers, and then increase again. This latter increase continues until the first node of the zeroth-order Shapiro step appears and after which the diode efficiency shows an oscillatory behavior.

We now investigate the microwave response of SQUID-1 at an even higher microwave frequency. Figures 3d and 3e show Shapiro maps of SQUID-1 measured at microwave frequency $f \sim 9.2$ GHz for $\Phi \sim \Phi_-$ and $\Phi \sim \Phi_+$, respectively. Integer and half-integer Shapiro steps are also observed in these measurements (see again a few labelled low-order step regions in the two figures). Additional data, including Shapiro maps measured at $\Phi \sim 0$ and $\Phi \sim 0.5\,\Phi_0$, and the magnetic-flux evolution of the Shapiro steps, are presented in Supplementary Section 1 (Figs. S2 and S3). Figure 3f presents the corresponding diode efficiency $\eta$ extracted as a function of power $P$ from Fig. 3d and 3e. As the microwave power increases, $\eta$ initially exhibits a weak power dependence, and then drops to zero and undergoes a sign-reversal with further increasing $P$, followed by oscillations at higher powers. Similar sign-reversal behavior of the diode efficiency is also observed at two higher frequencies, $f \sim 9.4$ GHz and $\sim 10.8$ GHz (see Supplementary

Section 1 Fig. S4 and Fig. S5).

The observed departure from the adiabatic behavior at these higher frequencies indicates that the quasistatic description is no longer sufficient to determine the edges of the zeroth Shapiro step. For instance, within the RSJ framework, when the device is driven beyond the adiabatic regime, even if the instantaneous bias current temporarily exceeds one of the switching thresholds in the absence of microwave irradiation, the time spent beyond the threshold may be too short for the superconducting phase to complete a phase slip before the microwave drive reverses. The edges of the zeroth step must instead be determined by solving the full time-dependent phase dynamics governed by Eq. (1). Consequently, the detailed form of the CPR becomes important in determining how the microwave drive modifies the positive and negative switching currents. We therefore extract the CPRs of the two junctions by fitting the switching currents obtained from the Shapiro maps [41], as shown in Figs. 3d and 3e and Supplementary Figs. S2a and S2b, measured at the microwave frequency of $f \sim 9.2$ GHz but the four different flux values of $\Phi \sim \Phi_-$, $\Phi_+$, 0, and $0.5\Phi_0$. In the theory fits, only the first two harmonics in each junction CPR are considered, which is reasonable because only the half-integer Shapiro steps are clearly observed in the measurements. The fits yield coefficients $a_1^1 = 1.043$ μA and $a_2^1 = -0.435$ μA for JJ1 and coefficients $a_1^2 = 0.452$ μA and $a_2^2 = -0.057$ μA for JJ2 in SQUID-1. Here $a_m^j$ denotes the coefficient of the $m$th harmonic of junction $j$ (for further details and the results of the fits, see Methods and Supplementary Section 3). We would like to note that these coefficients of the two JJs can also be extracted by fits to the switching currents shown in Fig. 1e measured for SQUID-1 without microwave irradiation. We have performed such fits (see the fitting details and results in Supplementary Section 3), which yields coefficients $a_1^1 = 1.006$ μA and $a_2^1 = -0.504$ for JJ1 and coefficients $a_1^2 = 0.411$ μA and $a_2^2 = -0.064$ μA for JJ2. It is evident that the agreement between the two sets of coefficients independently extracted by the two procedures is excellent, showing that our extracted CPRs for the two JJs in SQUID-1 are reliable. It is now worthwhile to note that our extracted coefficients for JJ1 are larger in magnitude than their corresponding values for JJ2, which is consistent with the fact that JJ1 has a wider transport channel when compared to JJ2. It is also worth noting that the second-harmonic components in the CPRs of both junctions are smaller than their corresponding first-harmonic components, as expected, while the magnitude ratio between the second- and first-harmonic components in JJ1 reaches approximately 0.42. This sizable second-harmonic contribution plays an important role in producing the pronounced SDE observed here and could also be advantageous for realizing parity-protected qubits [15–17, 20, 39, 40].

We now turn to numerical simulations to analyze the microwave driven dynamics of SQUID-1 within the RSJ model described by Eq. (1). In the simulations, the driving frequency is characterized by the reduced frequency $\Omega = f/f_J$, where $f$ is the microwave frequency and $f_J$ is the characteristic Josephson frequency of the device. We have performed simulations over a broad range of reduced frequencies, from $\Omega = 0.1$ to $10$, covering the evolution from adiabatic to strongly nonadiabatic phase dynamics. The simulation results shown in Fig. 4 are obtained for $\Omega = 0.1$, $0.6$, and $0.8$, which are selected to match the experimental measurements in Figs. 2 and 3 with a single assumed $f_J \sim 12\,\text{GHz}$. To compare with the experimental results, we use a dimensionless drive amplitude, given by $P = 20\log_{10}(I_{\text{ac}}) + P_0$, where $P_0$ is the offset used to align with experimental data. The simulated Shapiro maps shown in Figs. 4a, 4c, and 4e are obtained at $\Phi \sim \Phi_+$, while the corresponding results at $\Phi \sim \Phi_-$ are presented in Supplementary Section 3 (Fig. S11). Overall, the simulated Shapiro maps capture the main features of those observed experimentally. Moreover, the key features of the diode efficiency as a function of microwave power, namely the monotonic increase toward an ideal diode effect for $f \sim 1.2\,\text{GHz}$, the non-monotonic tuning of the diode efficiency for $f \sim 7.0\,\text{GHz}$, and the sign-reversal of the diode efficiency for $f \sim 9.2$ GHz, are all well captured by the simulated results shown in Figs. 4b, 4d, and 4f. For the simulated $\eta(P)$ beyond the adiabatic limit, as shown in Figs. 4d and 4f, deviations from the experimental results are observed. In this regime, heating is a possible effect that can modify the CPRs of the two junctions and is not included in the model [37, 41]. Nonequilibrium processes may also be induced by microwave irradiation and can renormalize the CPRs of the two junctions [26, 41, 42]. However, we do not have direct evidence to confirm their role in our measurements. Note that these deviations are most pronounced near the first node of the zeroth Shapiro step (indicated by black arrows in Figs. 4d and 4f), where the switching current becomes very small and the diode efficiency becomes highly sensitive to small variations in its value. Nevertheless, the simulations successfully reproduce the main experimental trends, indicating that the RSJ framework captures the dominant mechanism underlying the microwave tuning of the SDE in our device. Additional simulation results at higher frequencies, from $\Omega = 1$ to $\Omega = 10$, are provided in Supplementary Section 3 (Figs. S12 and S13). For these cases, the polarity reversal persists, while the corresponding power at which it occurs shifts to higher values as the frequency increases. This trend is also observed experimentally for $f = 9.2\,\text{GHz}$, $9.4\,\text{GHz}$, and $10.8\,\text{GHz}$, where the power required for polarity reversal increases with increasing frequency (see Fig. 3f, Fig. S4 and Fig. S5).

# 3 Discussion

We have fabricated SQUIDs with each having two bilayer graphene JJs embedded in an Al superconducting loop and found that the devices exhibit a pronounced SDE. We demonstrate that the pronounced SDE observed initially in the absence of microwave irradiation can be attributed to the combination of device asymmetry (e.g., the structural difference in the two bilayer-graphene JJs) and sizable second-harmonic components in the CPRs of the two JJs. For example, in JJ1 of SQUID-1, the ratio in magnitude between the second- and first-harmonic reaches approximately 0.42, which is relatively large compared with values reported for other SQUIDs exhibiting the SDE [15–17, 20]. From an application perspective, by employing individual gates on different bilayer graphene JJs, the asymmetry of the SQUIDs can be tuned more precisely [13, 14]. Such control could allow a bilayer-graphene SQUID to be tuned to a regime in which the first-harmonic contributions from the two JJs cancel each other, offering not only a possibility to further enhancing the SDE but also a compelling opportunity to realize parity-protected qubits [17, 39, 40]. Moreover, since the maximum diode efficiency grows as $(N_j - 1)/N_j$, where $N_j$ is the number of JJs [13], extending a double-junction interferometer to include additional JJs offers another route for enhancing the SDE. This approach is particularly feasible for graphene-based devices due to its two-dimensional nature. A larger initial diode efficiency could enable the unity diode efficiency to be reached at a lower microwave power under an adiabatic microwave driving. Such an operating condition could be advantageous for microwave controlled superconducting diode applications.

In conclusion, we have demonstrated the SDE in bilayer-graphene-based SQUIDs and studied how the SDE in the devices responds to microwave power at different driving frequencies. We have found that the diode efficiency of our SQUIDs can be effectively tuned by the microwave power and the tuning characteristics exhibit a pronounced frequency dependence. In particular, at sufficiently high driving frequencies, the diode efficiency exhibits a non-monotonic power dependence, including a sign reversal followed by oscillations at higher powers. We have analyzed these experimental results using simulations based on the RSJ model and found that main features of the observed frequency-dependent response of the SDE can be well captured by the simulations. These results demonstrate the potential of bilayer-graphene Josephson interferometers as a tunable platform for studying microwave-controlled superconducting diode physics, in which both the magnitude and polarity of the SDE can be controlled by microwave irradiation power. Our work could motivate further studies of microwave-driven superconductivity, light–matter interactions, and dynamic responses of topological states in bilayer-graphene-based hybrid devices [24, 43, 44].

# 4 Methods

## 4.1 Device Fabrication

Bulk hBN and graphite were purchased commercially (hBN: HQ-Graphene, Netherlands; graphite: NGS Naturgraphit GmbH, Germany). hBN/BLG/hBN van der Waals heterostructures were assembled using the dry transfer method similar to that reported in Ref. [31], in an Ar-filled glove box. Standard electron beam lithography (EBL) was used to define the contact regions. Immediately after selectively etching the top hBN layer in the contact regions by reactive ion etching (RIE) with $CF_4$ (60 sccm, 10 W) [45], superconducting contact electrodes of Ti/Al (4/30 nm) were deposited by thermal evaporation followed by a standard lift-off process.

## 4.2 Transport Measurements

Quasi-four-terminal measurements were conducted using a DC measurement technique in a dilution refrigerator equipped with a uniaxial magnet (Triton XL, Oxford). Specifically, a DC current bias was applied by a voltage generator (QDAC II, Quantum Machines) through a constant resistor (1 MΩ). The DC voltages were amplified by a voltage preamplifier (SR560, Stanford Research Systems) at room temperature before being measured by a multimeter (34461A, Keysight). The current applied to the magnet was supplied by a voltage/current DC source (GS210, Yokogawa). Microwave power was applied using a signal generator (E8267D, Keysight), with a 29 dBm attenuation designed in the dilution refrigerator. Microwave irradiation was introduced via an antenna positioned above the sample. Due to the unknown coupling between the antenna and the device, the exact power applied to the device is difficult to estimate. Data are presented in terms of the output power of the microwave generator, which is a common practice in such experiments.

## 4.3 Theoretical Analysis

For a SQUID device, the total current $I(\varphi)$ passing through is the sum of the currents from JJ1 ($I^1(\varphi)$) and JJ2 ($I^2(\varphi + \delta)$):

$$I(\varphi) = I^1(\varphi) + I^2(\varphi + \delta),$$

Where $\delta = 2\pi\Phi/\Phi_0$, with $\Phi$ being the applied magnetic flux and $\Phi_0 = h/2e$ the flux quantum. The CPR of each junction can be written as a Fourier series expansion:

$$I^j(\varphi) = \sum_{m=1}^{N} a_m^j \sin\left(m\varphi + \gamma_m^j\right).$$

Here, coefficients $a_m^j$ denotes the coefficient of the $m$th harmonics in the CPR of junction $j$, with $j$ = 1 and 2 referring to JJ1 and JJ2, respectively. The phase shift of the $m$th harmonic of junction $j$ is denoted by $\gamma_m^j$. With a similar method as described in Ref. [41], we simultaneously fitted the measured switching currents $I_{sw}^+$ and $|I_{sw}^-|$ of the four Shapiro maps shown in Figs. 3d and 3e and Fig. S2 using a single set of parameters. The fitting was performed over a selected range of microwave powers ($P$ < 6 dBm) to minimize the effects of heating and potential nonequilibrium processes at higher power. The fitting results are presented in Supplementary Section 3 (Fig. S9). The effect of inductance is neglected due to its relatively small value (see Supplementary Section 4) and the fact that the data can be well fitted without considering it.

## Acknowledgements

The authors would like to thank Dr. Xinjian Wei, Dr. Chuanwu Cao and Dr. Jianhuan Wang for helpful discussions on device fabrication techniques. This work was financially supported by the Beijing Natural Science Foundation (Grant No. 1264079, Grant No. JQ26005), the National Natural Science Foundation of China (Grant Nos. 92565304, 12374480, and 92576112), the National Key Research and Development Program of China (Grant No.2025YFE0217400), the Spanish Comunidad de Madrid (CM) "Talento Program" (Project 14 No. 2022-T1/IND-24070), the Spanish Ministry of Science, Innovation, and Universities through Grants CEX2024-001445-S (Severo Ochoa Centres of Excellence program) and PID2022-140552NA-I00. J.D. was supported by the Research Council of Norway through Grant No. 353894.

## Competing interests

The authors declare no conflict of interests.

## Data availability

The raw data and plotting code supporting the findings of this study are available upon reasonable request.

## Author contribution

H.Q.X. supervised the project. S.Y. fabricated the devices with help from H.S., J.Z., H.G., and X.W.. S.Y. performed the transport measurements with help from J.-Y.W. and Y.L.. R.S.S. and J.D. performed the theoretical analysis. S.Y., J.-Y.W., R.S.S., J.D., and H.Q.X. analyzed the results. S.Y., R.S.S., and H.Q.X. wrote the manuscript with inputs from all the authors.

## References

[1] Ando, F., Miyasaka, Y., Li, T., Ishizuka, J., Arakawa, T., Shiota, Y., Moriyama, T., Yanase, Y. & Ono, T. Observation of superconducting diode effect. *Nature* **584**, 373–376 (2020).

[2] Nadeem, M., Fuhrer, M. S. & Wang, X. The superconducting diode effect. *Nat. Rev. Phys.* **5**, 558–577 (2023).

[3] Ma, J., Zhan, R. & Lin, X. Superconducting diode effects: Mechanisms, materials and applications. *Adv. Phys. Res.* **4**, 2400180 (2025).

[4] Ghosh, S., Patil, V., Basu, A., Kuldeep, Dutta, A., Jangade, D. A., Kulkarni, R., Thamizhavel, A., Steiner, J. F., Oppen, Deshmukh, M. M. High-temperature Josephson diode. *Nat. Mater.* **23**, 612–618 (2024).

[5] Qi, S., Ge, J., Ji, C., Ai, Y., Ma, G., Wang, Z., Cui, Z., Liu, Y., Wang, Z. & Wang, J. High-temperature field-free superconducting diode effect in high-$T_c$ cuprates. *Nat. Commun.* **16**, 531 (2025).

[6] Wang, H., Zhu, Y., Bai, Z., Lyu, Z., Yang, J., Zhao, L., Zhou, X., Xue, Q.-K. & Zhang, D. Quantum superconducting diode effect with perfect efficiency above liquid-nitrogen temperature. *Nat. Phys.* **21**, 1–7 (2025).

[7] Mazur, G. P., Loo, N., Driel, D., Wang, J.-Y., Kouwenhoven, L. P., Badawy, G., Gazibegovic, S. & Bakkers, E. Gate-tunable Josephson diode. *Phys. Rev. Appl.* **22**, 054034 (2024).

[8] Su, H., Wang, J.-Y., Gao, H., Luo, Y., Yan, S., Wu, X., Li, G., Shen, J., Lu, L., Pan, D ., Zhao, J., Zhang, P., H.Q.X. Microwave-assisted unidirectional superconductivity in Al–InAs nanowire–Al junctions under magnetic fields. *Phys. Rev. Lett.* **133**, 087001 (2024).

[9] Yan, S., Luo, Y., Su, H., Gao, H., Wu, X., Pan, D., Zhao, J., Wang, J.-Y. & H.Q.X. Gate-tunable Josephson diode effect in Josephson junctions made from InAs nanosheets. *Adv. Funct. Mater.* **35**, 2503401 (2025).

[10] Turini, B., Salimian, S., Carrega, M., Iorio, A., Strambini, E., Giazotto, F., Zannier, V., Sorba, L. & Heun, S. Josephson diode effect in high-mobility InSb nanoflags. *Nano Lett.* **22**, 8502–8508 (2022).

[11] Lin, J.-X., Siriviboon, P., Scammell, H. D., Liu, S., Rhodes, D., Watanabe, K., Taniguchi, T., Hone, J., Scheurer, M. S. & Li, J. Zero-field superconducting diode effect in small-twist-angle trilayer graphene. *Nat. Phys.* **18**, 1221–1227 (2022).

[12] Diez-Merida, J., Díez-Carlón, A., Yang, S., Xie, Y.-M., Gao, X.-J., Senior, J., Watanabe, K., Taniguchi, T., Lu, X., Higginbotham, A. P., Law, K. T., Efetov, D. K. Symmetry-broken

Josephson junctions and superconducting diodes in magic-angle twisted bilayer graphene. *Nat. Commun.* **14**, 2396 (2023).

[13] Seoane Souto, R., Leijnse, M. & Schrade, C. Josephson diode effect in supercurrent interferometers. *Phys. Rev. Lett.* **129**, 267702 (2022).

[14] Ciaccia, C., Haller, R., Drachmann, A. C., Lindemann, T., Manfra, M. J., Schrade, C. & Schönenberger, C. Gate-tunable Josephson diode in proximitized InAs supercurrent interferometers. *Phys. Rev. Res.* **5**, 033131 (2023).

[15] Paolucci, F., De Simoni, G. & Giazotto, F. A gate- and flux-controlled supercurrent diode effect. *Appl. Phys. Lett.* **122**, 041601 (2023).

[16] Zhang, P., Zarassi, A., Jarjat, L., Sande, V., Pendharkar, M., Lee, J., Dempsey, C. P., McFadden, A., Harrington, S. D., Dong, J. T., Wu, H., Chen, A.-H., Hocevar, M., Palmstrøm, C. J., Frolov, S. M. Large second-order Josephson effect in planar superconductor–semiconductor junctions. *SciPost Phys.* **16**, 030 (2024).

[17] Valentini, M., Sagi, O., Baghumyan, L., Gijsel, T., Jung, J., Calcaterra, S., Ballabio, A., Aguilera Servin, J., Aggarwal, K., Janik, M, Adletzberger, T., Seoane Souto, R., Leijnse, M., Danon, J., Schrade, C., Bakkers, E. P. A. M., Chrastina, D., Isella, G., Katsaros, G. Parity-conserving Cooper-pair transport and ideal superconducting diode in planar germanium. *Nat. Commun.* **15**, 169 (2024).

[18] Cuozzo, J. J., Pan, W., Shabani, J. & Rossi, E. Microwave-tunable diode effect in asymmetric SQUIDs with topological Josephson junctions. *Phys. Rev. Res.* **6**, 023011 (2024).

[19] Leblanc, A., Tangchingchai, C., Momtaz, Z. S., Kiyooka, E., Hartmann, J.-M., Fernandez-Bada, G. T., Scherübl, Z., Brun, B., Schmitt, V., Zihlmann, S, Maurand, R., Dumur, É., De Franceschi, S., Lefloch, F. From nonreciprocal to charge-4e supercurrent in Ge-based Josephson devices with tunable harmonic content. *Phys. Rev. Res.* **6**, 033281 (2024).

[20] Wu, X., Wang, J.-Y., Su, H., Yan, S., Pan, D., Zhao, J., Zhang, P. & H.Q.X. Tunable superconducting diode effect in higher-harmonic InSb nanosheet interferometers. *New J. Phys.* **27**, 023031 (2025).

[21] Shapiro, S. Josephson currents in superconducting tunneling: The effect of microwaves and other observations. *Phys. Rev. Lett.* **11**, 80 (1963).

[22] Rokhinson, L. P., Liu, X. & Furdyna, J. K. The fractional ac Josephson effect in a semiconductor–superconductor nanowire as a signature of Majorana particles. *Nat. Phys.* **8**, 795–799 (2012).

[23] Seoane Souto, R., Leijnse, M., Schrade, C., Valentini, M., Katsaros, G. & Danon, J. Tuning the Josephson diode response with an ac current. *Phys. Rev. Res.* **6**, 022002 (2024).

[24] Park, S., Lee, W., Jang, S., Choi, Y.-B., Park, J., Jung, W., Watanabe, K.,Taniguchi, T., Cho, G.Y., Lee, G.-H. Steady Floquet–Andreev states in graphene Josephson junctions. *Nature* **603**(7901), 421–426 (2022).

[25] Haxell, D. Z., Coraiola, M., Sabonis, D., Hinderling, M., Ten Kate, S. C., Cheah, E., Krizek, F., Schott, R., Wegscheider, W., Belzig, W, Cuevas, J. C., Nichele, F.Microwave-induced conductance replicas in hybrid Josephson junctions without Floquet–Andreev states. *Nat. Commun.* **14**, 6798 (2023).

[26] Iorio, A., Crippa, A., Turini, B., Salimian, S., Carrega, M., Chirolli, L., Zannier, V., Sorba, L., Strambini, E., Giazotto, F., Heun, S. Half-integer Shapiro steps in highly transmissive InSb nanoflag Josephson junctions. *Phys. Rev. Res.* **5**, 033015 (2023).

[27] He, J., Pan, D., Liu, M., Lyu, Z., Jia, Z., Yang, G., Zhu, S., Liu, G., Shen, J., Shevchenko, S. N, Nori, F., Zhao, J., Lu, L., Qu, F. Quantifying quantum coherence of multiple-charge states in tunable Josephson junctions. *npj Quantum Inf.* **10**, 1 (2024).

[28] Matsuo, S., Deacon, R. S., Kobayashi, S., Sato, Y., Yokoyama, T., Lindemann, T., Gronin, S., Gardner, G. C., Ishibashi, K., Manfra, M. J, Tarucha, S. Shapiro response of superconducting diode effect derived from Andreev molecules. *Phys. Rev. B* **111**, 094512 (2025).

[29] Borgongino, L., Seoane Souto, R., Paghi, A., Senesi, G., Skibinska, K., Sorba, L., Riccardi, E., Giazotto, F. & Strambini, E. Biharmonic-drive tunable Josephson diode. *Nano Lett.* **25**, 14451–14458 (2025).

[30] Bhattacharyya, B., de Wit, S. R., Wu, Z., Huang, Y., Golden, M. S., Brinkman, A. & Li, C. Coexisting topological hinges and 1D Rashba states in $Bi_{0.97}Sb_{0.03}$ revealed by the Josephson effect. *arXiv* **:2505.02995** (2025).

[31] Wang, L., Meric, I., Huang, P., Gao, Q., Gao, Y., Tran, H., Taniguchi, T., Watanabe, K., Campos, L., Muller, D. A., Guo, J., Kim, P., Hone, J., Shepard, K. L., Dean, C. R. One-dimensional electrical contact to a two-dimensional material. *Science* **342**, 614–617 (2013).

[32] Sun, S. & Jarillo-Herrero, P. Optimized fabrication procedure for high-quality graphene-based moiré superlattice devices. *arXiv:* **2507.15853** (2025).

[33] Novoselov, K. S., Geim, A. K., Morozov, S. V., Jiang, D., Katsnelson, M. I., Grigorieva, I. V., Dubonos, S. V. & Firsov, A. A. Two-dimensional gas of massless Dirac fermions in graphene. *Nature* **438**, 197–200 (2005).

[34] Zhang, Y., Tan, Y.-W., Stormer, H. L. & Kim, P. Experimental observation of the quantum Hall effect and Berry's phase in graphene. *Nature* **438**, 201–204 (2005).

[35] Park, G.-H., Lee, W., Park, S., Watanabe, K., Taniguchi, T., Cho, G. Y. & Lee, G.-H.

Controllable Andreev bound states in bilayer graphene Josephson junctions from short to long junction limits. *Phys. Rev. Lett.* **132**, 226301 (2024).

[36] Sharma, A., Chen, C.-C., McCourt, J., Kim, M., Watanabe, K., Taniguchi, T., Rokhinson, L., Finkelstein, G. & Borzenets, I. Fermi velocity dependent critical current in ballistic bilayer graphene Josephson junctions. *ACS Nanoscience Au* **5**, 65–69 (2025).

[37] De Cecco, A., Le Calvez, K., Sacépé, B., Winkelmann, C. & Courtois, H. Interplay between electron overheating and ac Josephson effect. *Phys. Rev. B* **93**, 180505(R) (2016).

[38] Raes, B., Tubsrinuan, N., Sreedhar, R., Guala, D., Panghotra, R., Dausy, H., Souza Silva, C. C. & Vondel, J. Fractional Shapiro steps in resistively shunted Josephson junctions as a fingerprint of a skewed current–phase relationship. *Phys. Rev. B* **102**, 054507 (2020).

[39] Bell, M. T., Paramanandam, J., Ioffe, L. B. & Gershenson, M. E. Protected Josephson rhombus chains. *Phys. Rev. Lett.* **112**, 167001 (2014).

[40] Larsen, T. W., Gershenson, M. E., Casparis, L., Kringhøj, A., Pearson, N. J., McNeil, R. P. G., Kuemmeth, F., Krogstrup, P., Petersson, K. D. & Marcus, C. M. Parity-protected superconductor–semiconductor qubit. *Phys. Rev. Lett.* **125**, 056801 (2020).

[41] Tjøtta, M., Shah, D., Modi, K., Valentini, M., Seoane Souto, R., Katsaros, G. & Danon, J. Full Shapiro spectroscopy of current–phase relationships. *arXiv* **2511.14313** (2025).

[42] Dartiailh, M. C., Cuozzo, J. J., Elfeky, B. H., Mayer, W., Yuan, J., Wickramasinghe, K. S., Rossi, E. & Shabani, J. Missing Shapiro steps in topologically trivial Josephson junctions on InAs quantum wells. *Nat. Commun.* **12**, 78 (2021).

[43] Xie, Y.-M., Lantagne-Hurtubise, É., Young, A. F., Nadj-Perge, S. & Alicea, J. Gate-defined topological Josephson junctions in Bernal bilayer graphene. *Phys. Rev. Lett.* **131**, 146601 (2023).

[44] Peralta Gavensky, L., Usaj, G., Feinberg, D. & Balseiro, C. A. Berry curvature tomography and realization of the topological Haldane model in driven three-terminal Josephson junctions. *Phys. Rev. B* **97**, 220505(R) (2018).

[45] Jang, S., Park, G.-H., Park, S., Jeong, H.-W., Watanabe, K., Taniguchi, T. & Lee, G.-H. Engineering superconducting contacts transparent to a bipolar graphene. *Nano Lett.* **24**, 15582–15587 (2024)

# Figures

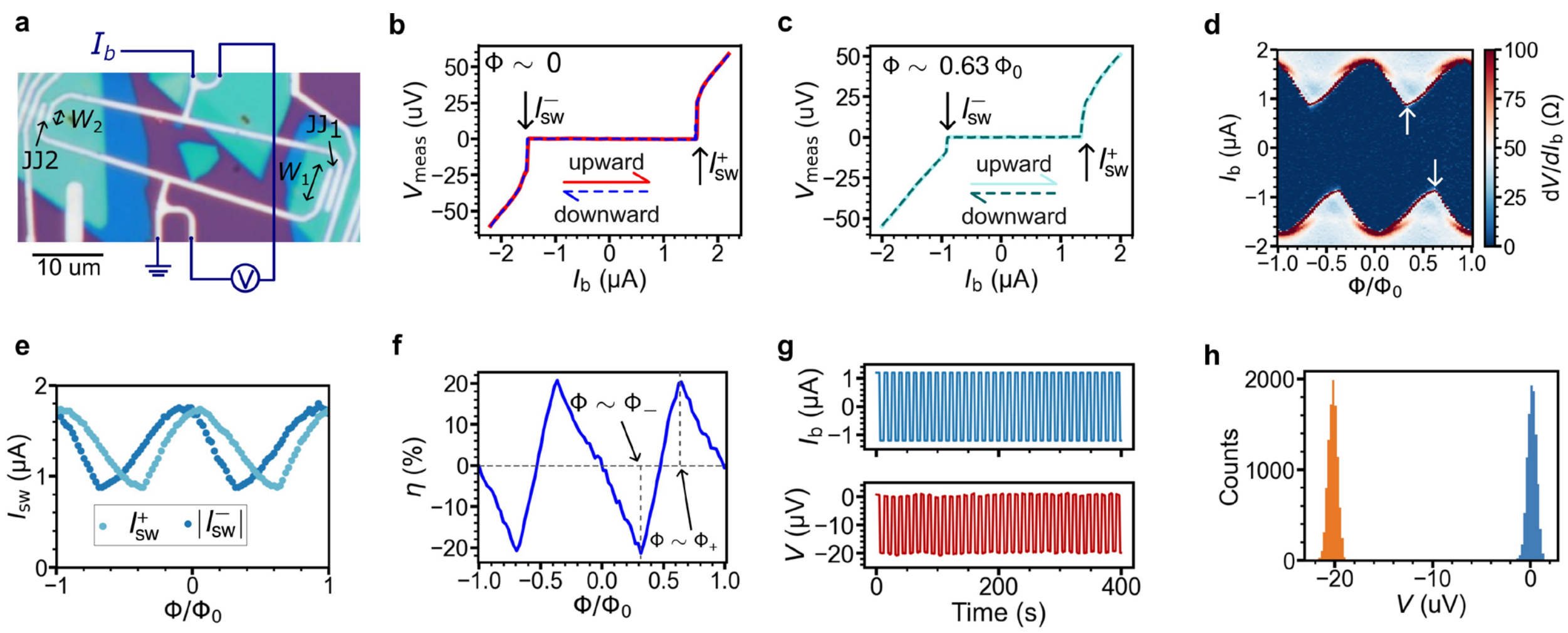

**Fig. 1 |Device and characterization of the SDE. a** Optical image of SQUID-1. The scale bar is 10 μm. The superconducting electrodes are made of 4/30-nm-thick Ti/Al bilayers. The two Josephson junctions of the SQUID are formed from two nearby hBN/BLG/hBN heterostructures. The transport measurements are conducted in a quasi-four-terminal circuit setup, as schematically illustrated in the figure. **b** Measured voltage $V_{\mathrm{meas}}$ of the SQUID as a function of bias current $I_{\mathrm{b}}$ at $V_{\mathrm{bg}} = 20\,\mathrm{V}$ and zero out-of-plane magnetic field (i.e., $\Phi \sim 0$). The red solid and blue dashed curves represent the measurement results during upward and downward current bias $I_{\mathrm{b}}$ sweeps, respectively. **c** Same as **b**, except with an out-of-plane magnetic field of $B \sim 6$ μT, corresponding to $\Phi \sim 0.63\,\Phi_0$. **d** Differential resistance $dV/dI_{\mathrm{b}}$ of the SQUID, as a function of normalized external flux $\Phi/\Phi_0$ and $I_{\mathrm{b}}$ (upward current sweeping direction only). The white arrows indicate the minima of $I_{\mathrm{sw}}^{+}$ and $|I_{\mathrm{sw}}^{-}|$. **e** Extracted $I_{\mathrm{sw}}^{+}$ and $|I_{\mathrm{sw}}^{-}|$ from **d**. **f** Diode efficiency $\eta$ as a function of $\Phi/\Phi_0$. The values of $\Phi = \Phi_{+}$ ($\sim 0.63\,\Phi_0$) and $\Phi = \Phi_{-}$ ($\sim 0.32\,\Phi_0$) represent the external magnetic fluxes at which $\eta$ reaches its maximum positive and negative values, respectively. **g** Rectification function of the SDE. The blue line in the upper panel shows a square-wave current with an amplitude of 1.2 μA, while the red line in the lower panel represents the voltage output measured across the SQUID. The measurements were performed for approximately 6.6 h, during which 20,000 voltage values were consecutively recorded. Here, in the figure, only a representative 400 s segment of the time trace is shown. **h** Histogram of the 20,000 consecutively recorded voltage values.

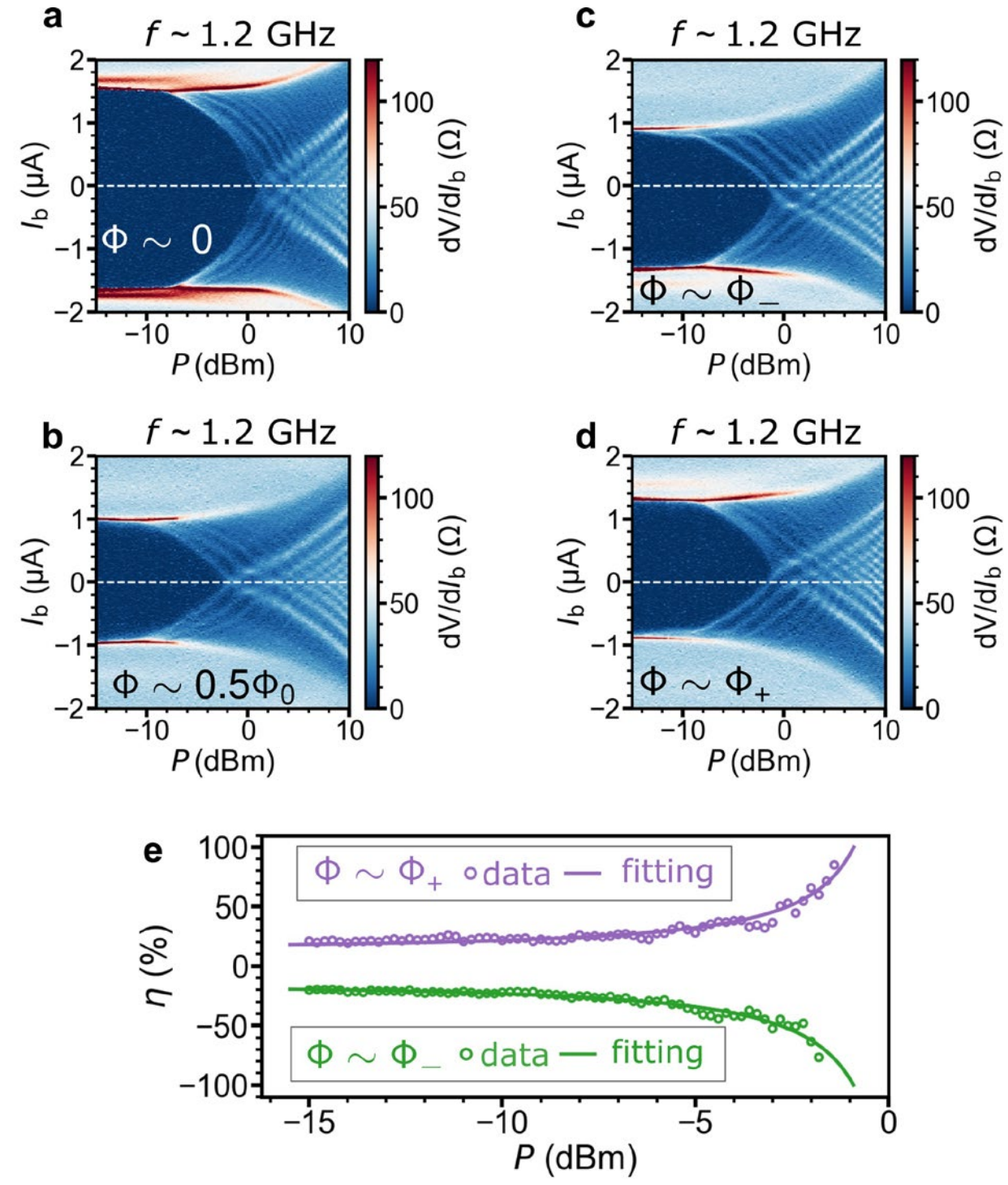

**Fig. 2 | Microwave response of the SDE under a low-frequency microwave irradiation. a–d** Differential resistance $dV/dI_{\mathrm{b}}$ of SQUID-1 as a function of radiation power $P$ and bias current $I_{\mathrm{b}}$, measured at $\Phi \sim 0,\ 0.5\,\Phi_0,\ \Phi_-$ and $\Phi_+$, respectively. The microwave frequency is set at 1.2 GHz. **e** Diode efficiency $\eta$ as a function of microwave power $P$. Green and purple circles represent the data extracted from the measurements at $\Phi \sim \Phi_-$ and $\Phi \sim \Phi_+$, respectively. Solid curves are the fitting results.

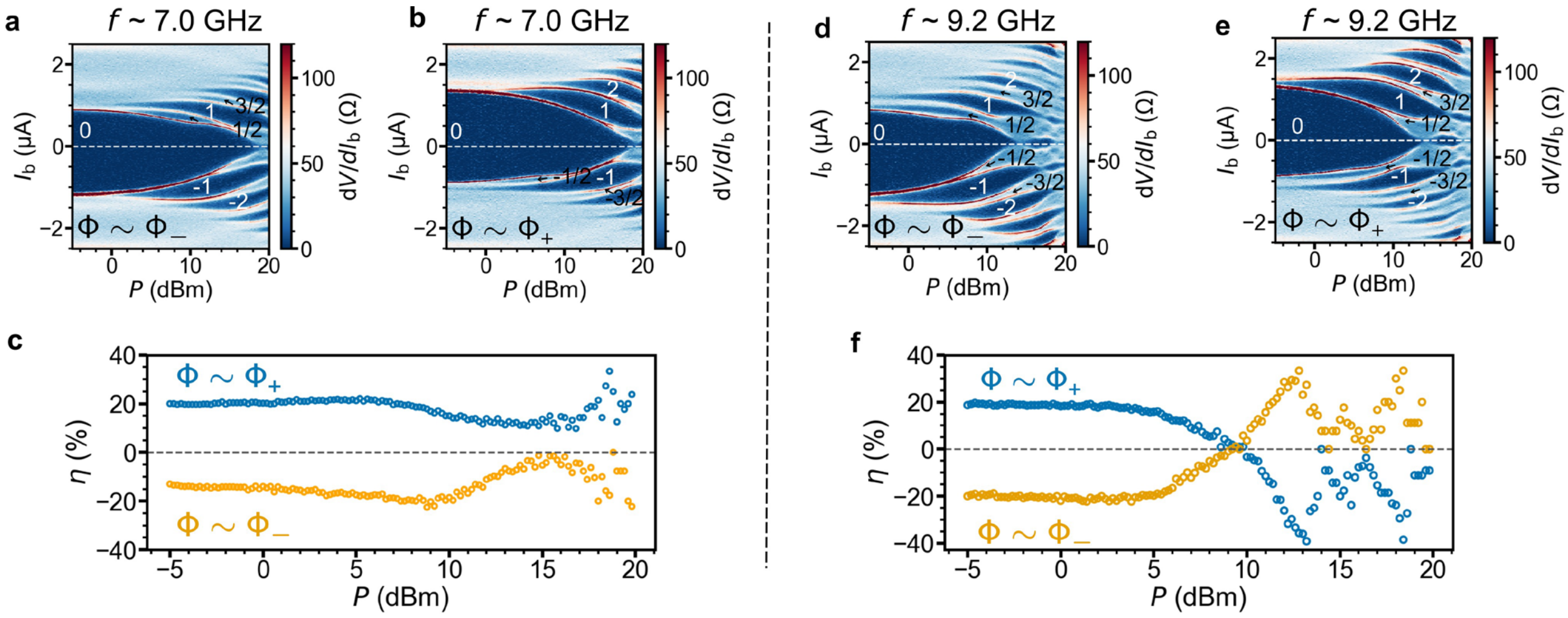

**Fig. 3 | Microwave response of the SDE under two high-frequency microwave irradiations. a, b** Differential resistance $dV/dI_b$ of SQUID-1 as a function of microwave power $P$ and bias current $I_b$, measured at $\Phi \sim \Phi_-$ and $\Phi \sim \Phi_+$, respectively. The microwave frequency is set at $f \sim$ 7.0 GHz. **c** Diode efficiency $\eta$ as a function of $P$ extracted from **a** and **b**, i.e., at $\Phi \sim \Phi_-$ (yellow circles) and $\Phi \sim \Phi_+$ (blue circles), respectively. **d, e** Differential resistance $dV/dI_b$ as a function of microwave power $P$ and bias current $I_b$, measured at $\Phi \sim \Phi_-$ and $\Phi \sim \Phi_+$, respectively. Here, the microwave frequency is set at $f \sim$ 9.2 GHz. **f** Diode efficiency $\eta$ as a function of $P$ extracted from **d** (yellow circles) and **e** (blue circles).

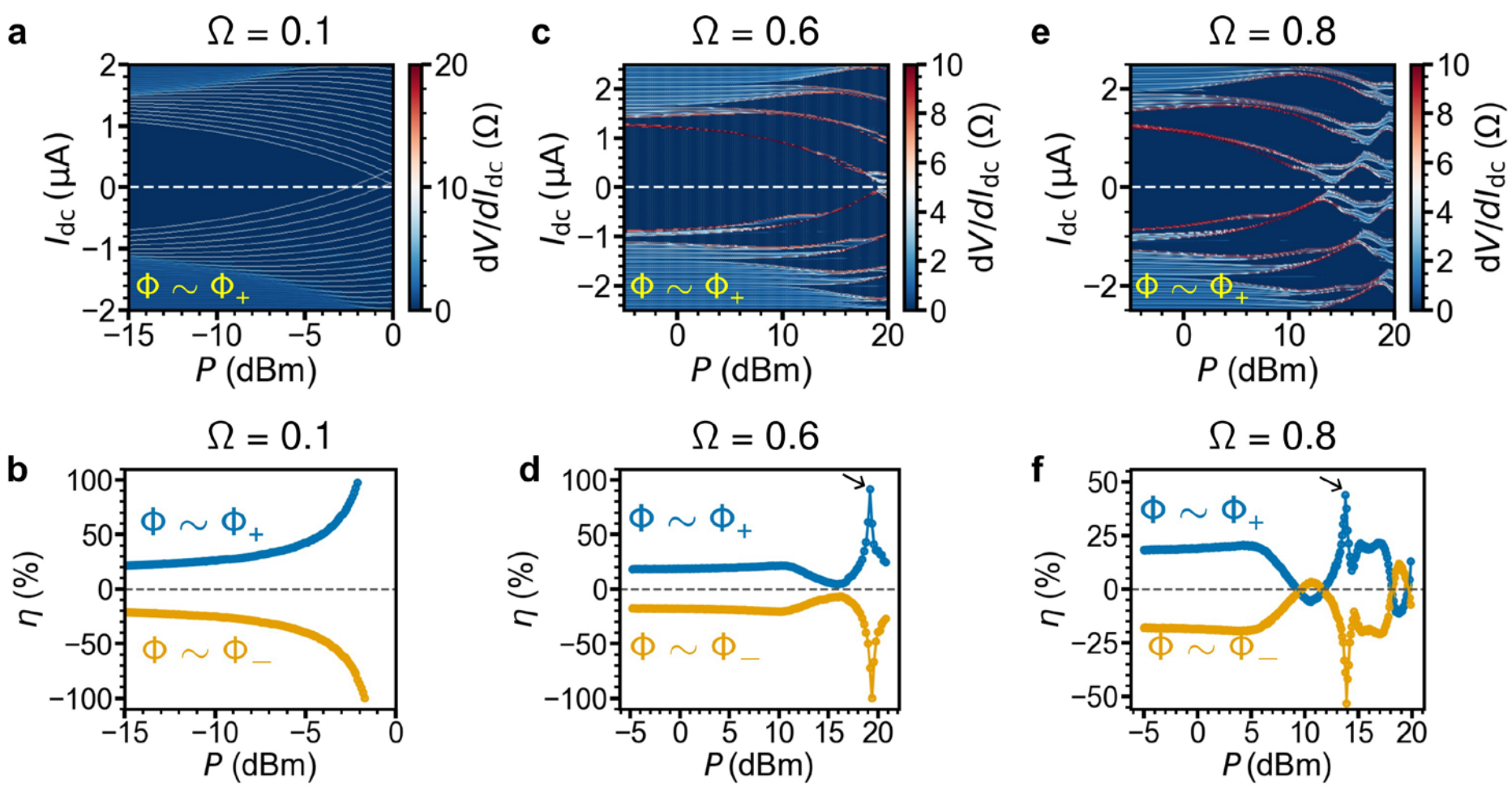

**Fig. 4 | Simulated microwave response of the SDE under microwave irradiations of different frequencies. a–f** Shapiro maps simulated using the extracted Josephson junction CPRs of SQUID-1 and diode efficiency $\eta$ of SQUID-1 obtained from the simulations as a function of microwave power $P$. The reduced frequencies are chosen as $\Omega = 0.1$, 0.6 and 0.8 for panels (**a**, **b**), (**c**, **d**) and (**e**, **f**), respectively. The black arrows in panels **d** and **f** mark the efficiency peaks extracted near the first node of the zeroth Shapiro step. Note that in panels **a**, **c**, and **e**, only the simulated results for $\Phi \sim \Phi_+$ are presented, while in panels **b**, **d**, and **f**, the simulated results for both $\Phi \sim \Phi_+$ and $\sim \Phi_-$ are presented.

# Supplementary Information: Microwave Response of the Superconducting Diode Effect in Proximitized Bilayer Graphene Interferometers

Shili Yan[1], Rubén Seoane Souto[2,3], Yi Luo[1], Jeroen Danon [4], Haitian Su[1], Junze Zhang[1], Han Gao[1], Xingjun Wu[1], Ji-Yin Wang[1], H.Q. Xu[1,5*]

[1] Beijing Academy of Quantum Information Sciences, Beijing, 100193, China.

[2] Instituto de Ciencia de Materiales de Madrid (ICMM), Consejo Superior de Investigaciones Científicas (CSIC), Sor Juana Inés de la Cruz 3, Madrid, 28049, Spain.

[3] Quantum Advanced Research Center (QuARC), Consejo Superior de Investigaciones Científicas (CSIC), Sor Juana Inés de la Cruz 3, Madrid, 28049, Spain.

[4] Department of Physics, Norwegian University of Science and Technology, Trondheim, NO-7491, Norway.

[5] Beijing Key Laboratory of Quantum Devices, Peking University, Beijing, 100871, China.

E-mail: hqxu@pku.edu.cn

## Section 1. Additional Data of SQUID-1

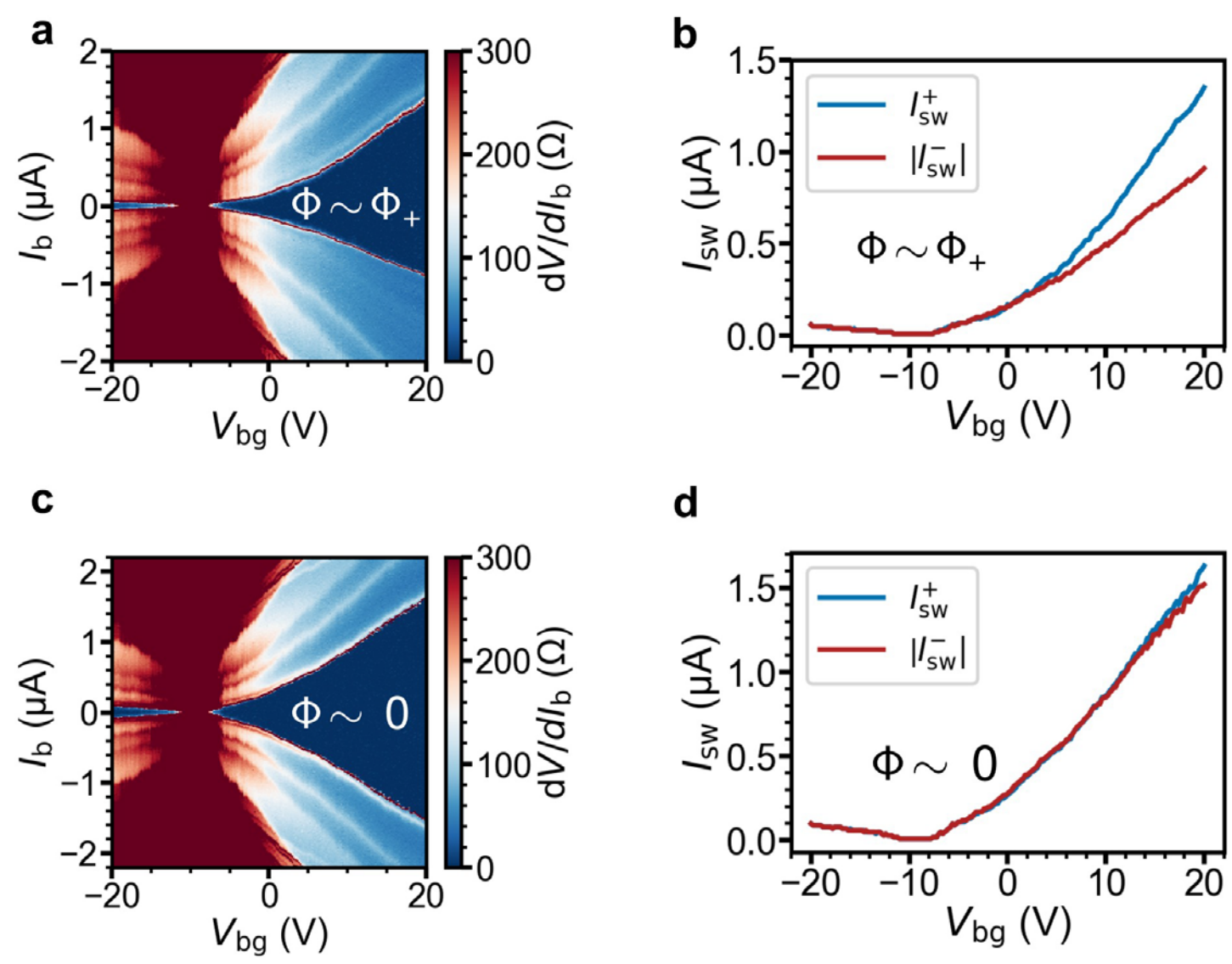

**Figure S1. Gate tunable SDE in SQUID-1. a** Differential resistance $dV/dI_b$ of SQUID-1 as a function of $V_{bg}$ and $I_b$ measured at $\Phi \sim \Phi_+$. **b** Extracted values of $I_{sw}^+$ and $|I_{sw}^-|$ from panel **a**. **c** Differential resistance $dV/dI_b$ of SQUID-1 as a function of $V_{bg}$ and $I_b$ measured at $\Phi \sim 0$. **d** Extracted values of $I_{sw}^+$ and $|I_{sw}^-|$ from panel **c**.

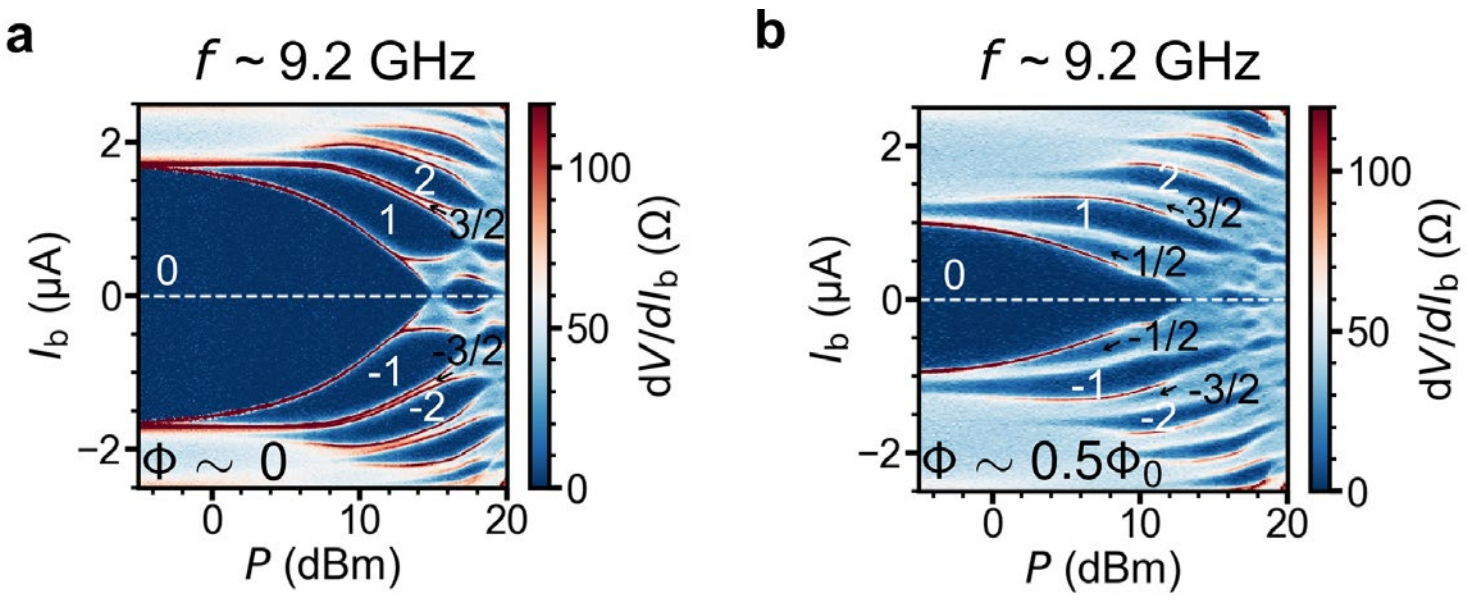

**Figure S2. a, b** Differential resistance $dV/dI_b$ of SQUID-1 as a function of microwave power $P$ and bias current $I_b$, measured at $\Phi \sim 0$ and $0.5\,\Phi_0$, respectively. The regions, in which a few low-order Shapiro steps are formed, are labelled by their corresponding step orders. Here, the microwave frequency is set at $f \sim 9.2$ GHz.

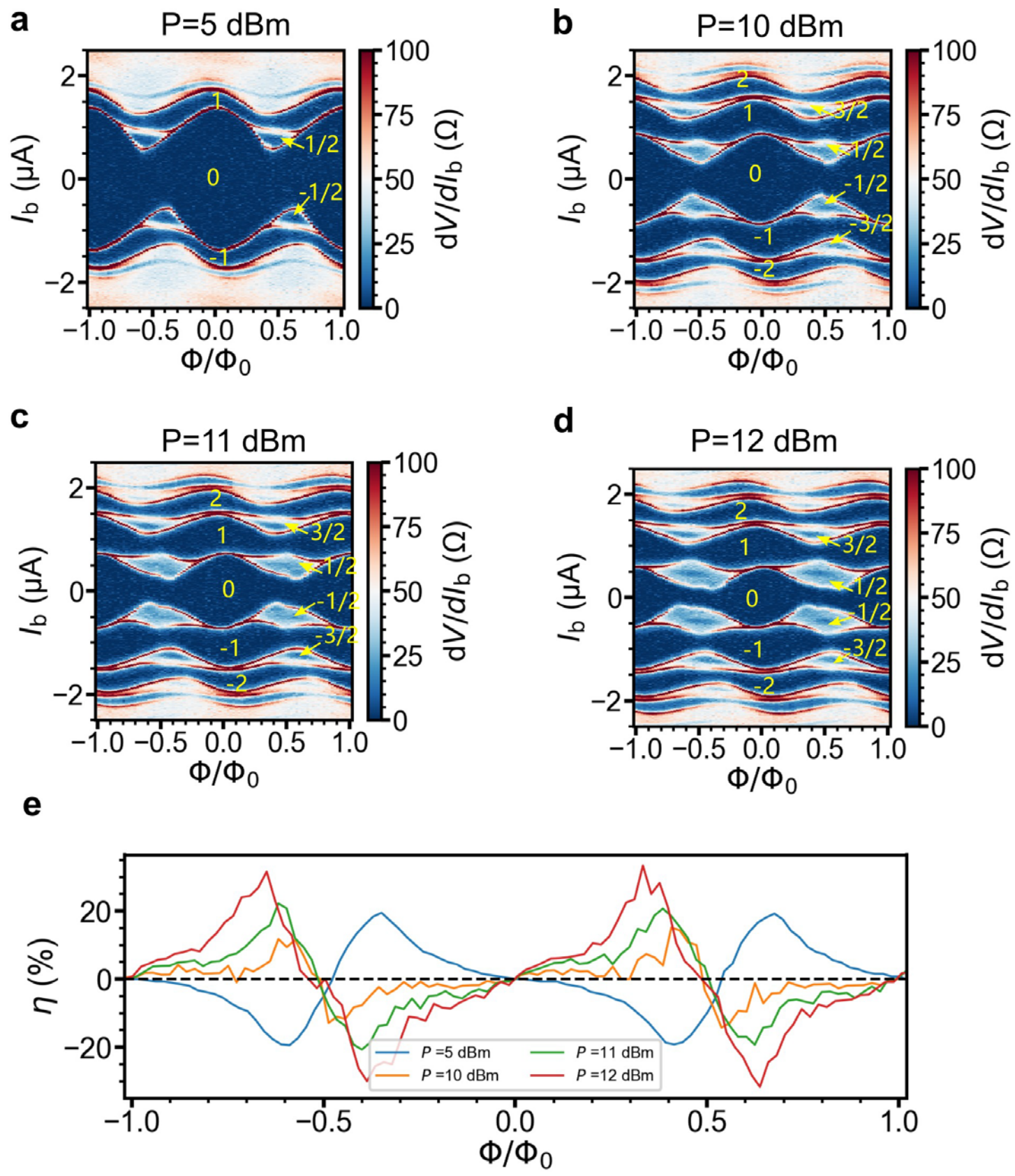

**Figure S3. a–d** Differential resistance $dV/dI_{\mathrm{b}}$ of SQUID-1 as a function of normalized magnetic flux $\Phi = \Phi/\Phi_0$ and bias current $I_{\mathrm{b}}$, measured under microwave irradiation with powers $P = 5$ dBm, 10 dBm, 11 dBm, and 12 dBm, respectively. **e** Diode efficiencies $\eta$, extracted from panels **a–d**, as a function of normalized magnetic flux $\Phi/\Phi_0$ at various microwave powers $P$. The microwave frequency is set at $f \sim 9.2$ GHz.

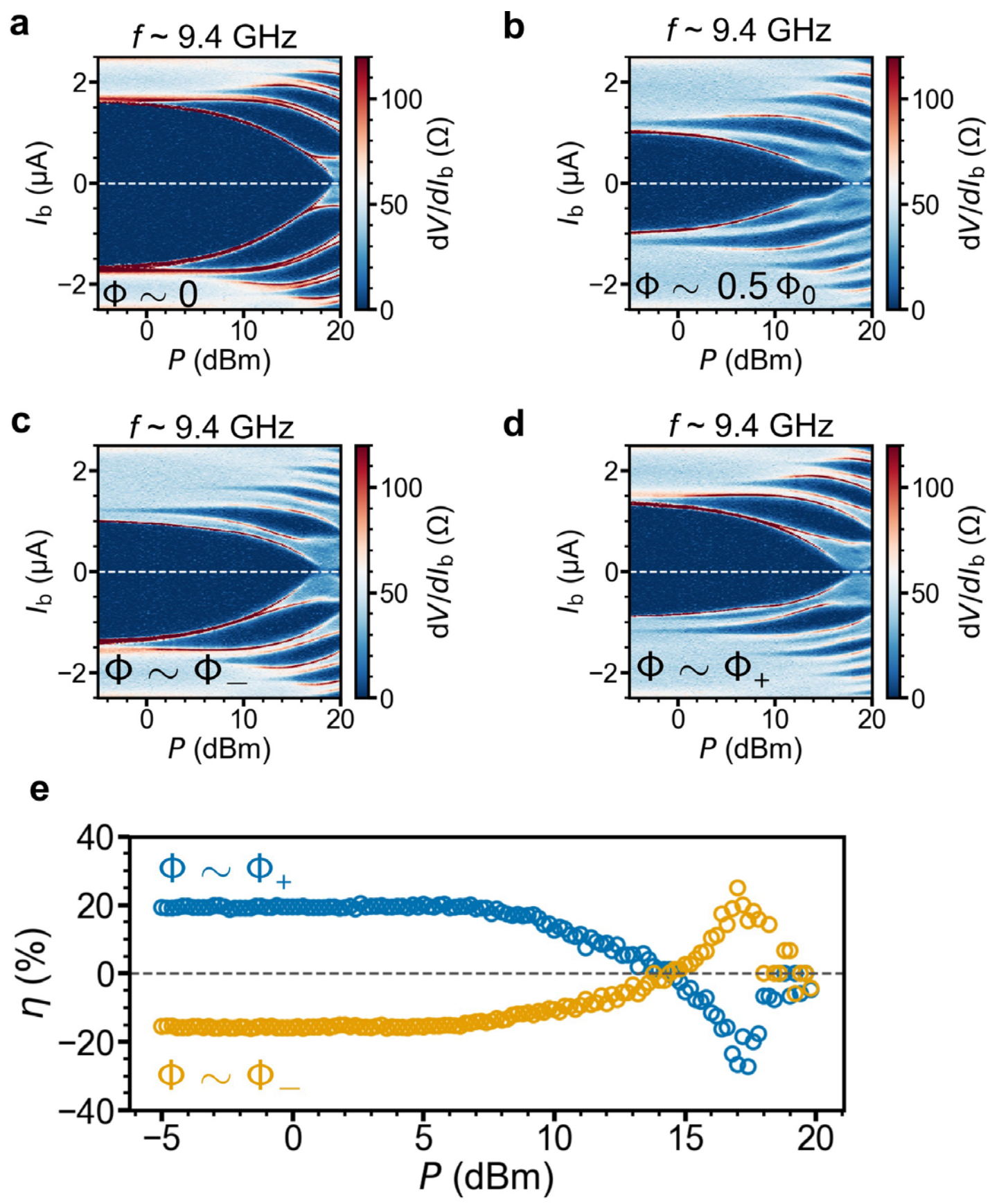

**Figure S4. Microwave response of the SDE under microwave irradiation with a frequency f∼ 9.4 GHz. a–d** Differential resistance $dV/dI_{\mathrm{b}}$ of SQUID-1 as a function of radiation power $P$ and bias current $I_{\mathrm{b}}$, measured at $\Phi \sim 0$, $0.5\,\Phi_0$, $\Phi_-$ and $\Phi_+$, respectively. The microwave frequency is set at $\sim$ 9.4 GHz. **e** Diode efficiencies $\eta$ as a function of $P$ extracted from **c** and **d**, i.e., at $\Phi \sim \Phi_-$ (yellow circles) and $\Phi \sim \Phi_+$ (blue circles), respectively.

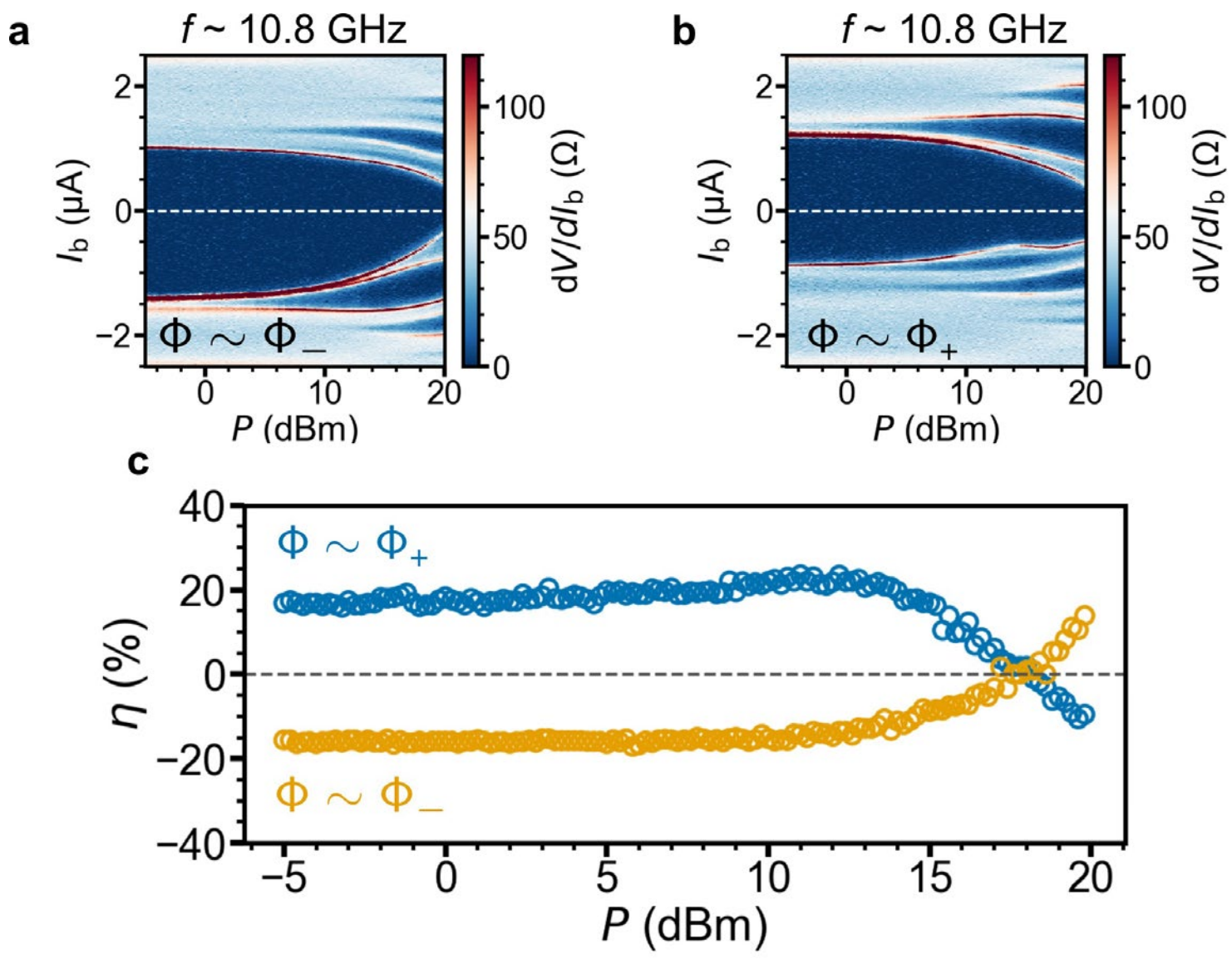

**Figure S5. Microwave response of the SDE under microwave irradiation with a frequency $\mathbf{f \sim 10.8}$ GHz. a–b** Differential resistance $dV/dI_\mathrm{b}$ of SQUID-1 as a function of radiation power $P$ and bias current $I_\mathrm{b}$, measured at $\Phi \sim 0$, $0.5\,\Phi_0$, $\Phi_-$ and $\Phi_+$, respectively. The microwave frequency is set at $\sim$ 10.8 GHz. **c** Diode efficiencies $\eta$ as a function of $P$ extracted from **a** and **b**, i.e., at $\Phi \sim \Phi_-$ (yellow circles) and $\Phi \sim \Phi_+$ (blue circles), respectively.

## Section 2. Additional Data of SQUID-2

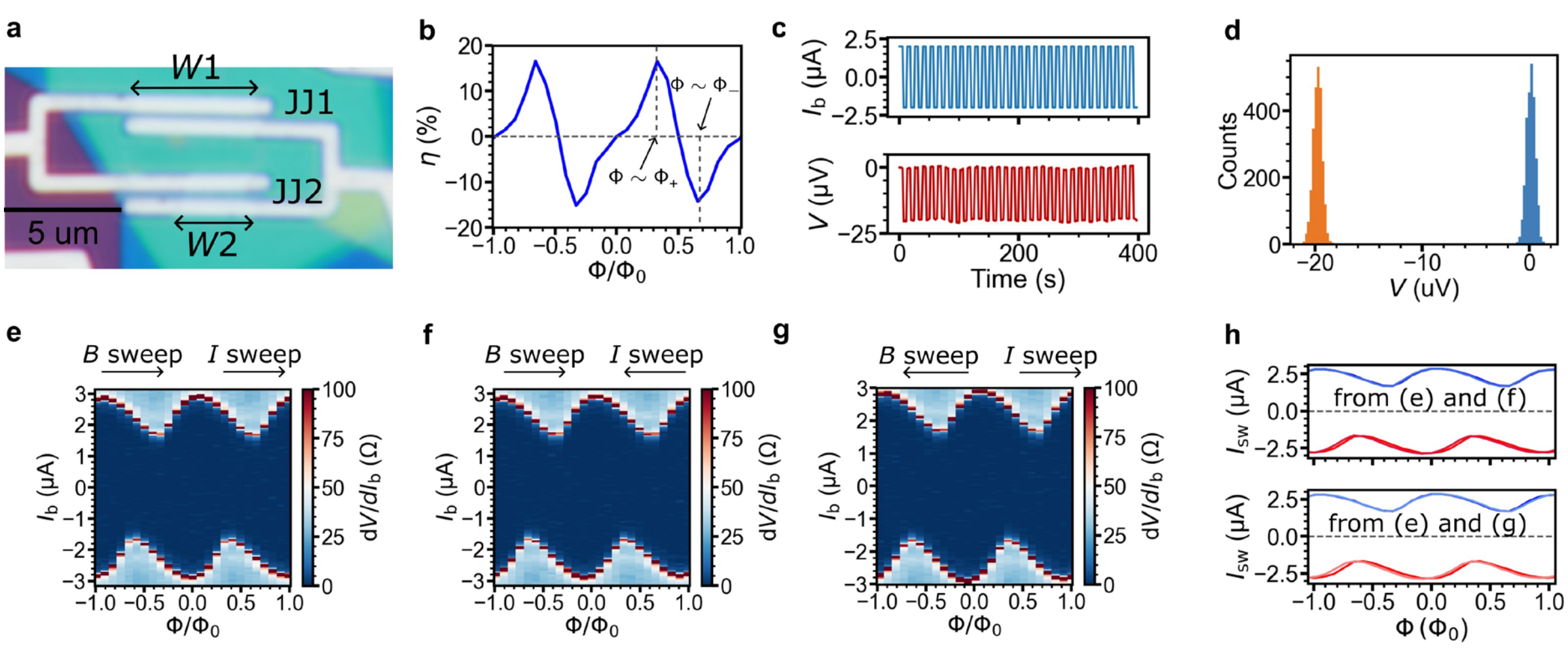

**Figure S6. Device structure of and characterization of the SDE in SQUID-2. a** Optical image of SQUID-2. The scale bar is 5 μm. The superconducting electrodes are made of 4/30-nm-thick Ti/Al bilayers. The two Josephson junctions (JJ1 and JJ2) of the SQUID are formed from one hBN/BLG/hBN heterostructure. In this SQUID, the junction JJ1 has a width of $W_1 = 5.4$ μm and a length (contact electrode separation) of $d_1 = 260$ nm, and the junction JJ2 has a width of $W_1 = 3.4$ μm and a length of $d_2 = 360$ nm. **b** Diode efficiency $\eta$ as a function of $\Phi/\Phi_0$ extracted from panel **e**. The values of $\Phi = \Phi_+$ and $\Phi = \Phi_-$ represent the external magnetic flux at which $\eta$ reaches its maximum positive and negative values, respectively. **c** Demonstration of diode rectification function. The blue trace in the upper panel shows a square-wave current with amplitude ~ 2 μA, while the red line in the lower panel represents the voltage output measured across the SQUID. The measurements were performed for approximately 11.7 h, during which 5,000 voltage values were consecutively recorded. Here, in the figure, only a representative 400 s segment of the time trace is shown. **d** Histogram of the 5,000 consecutively recorded voltage values. **e** Differential resistance $dV/dI_b$ of the SQUID, as a function of normalized applied flux $\Phi/\Phi_0$ and $I_b$, measured with upward magnetic-field and upward current sweeps. **f** Same as panel **e**, but measured with upward magnetic-field and downward current sweeps. **g** Same as panel **e**, but measured with downward sweeps of both magnetic field and current. **h** Upper panel: Extracted $I_{sw}^+$ and $|I_{sw}^-|$ from panels **e** and **f**, showing that the traces nearly overlap. Lower panel: Extracted $I_{sw}^+$ and $|I_{sw}^-|$ from panels **e** and **g**, showing that the traces nearly overlap.

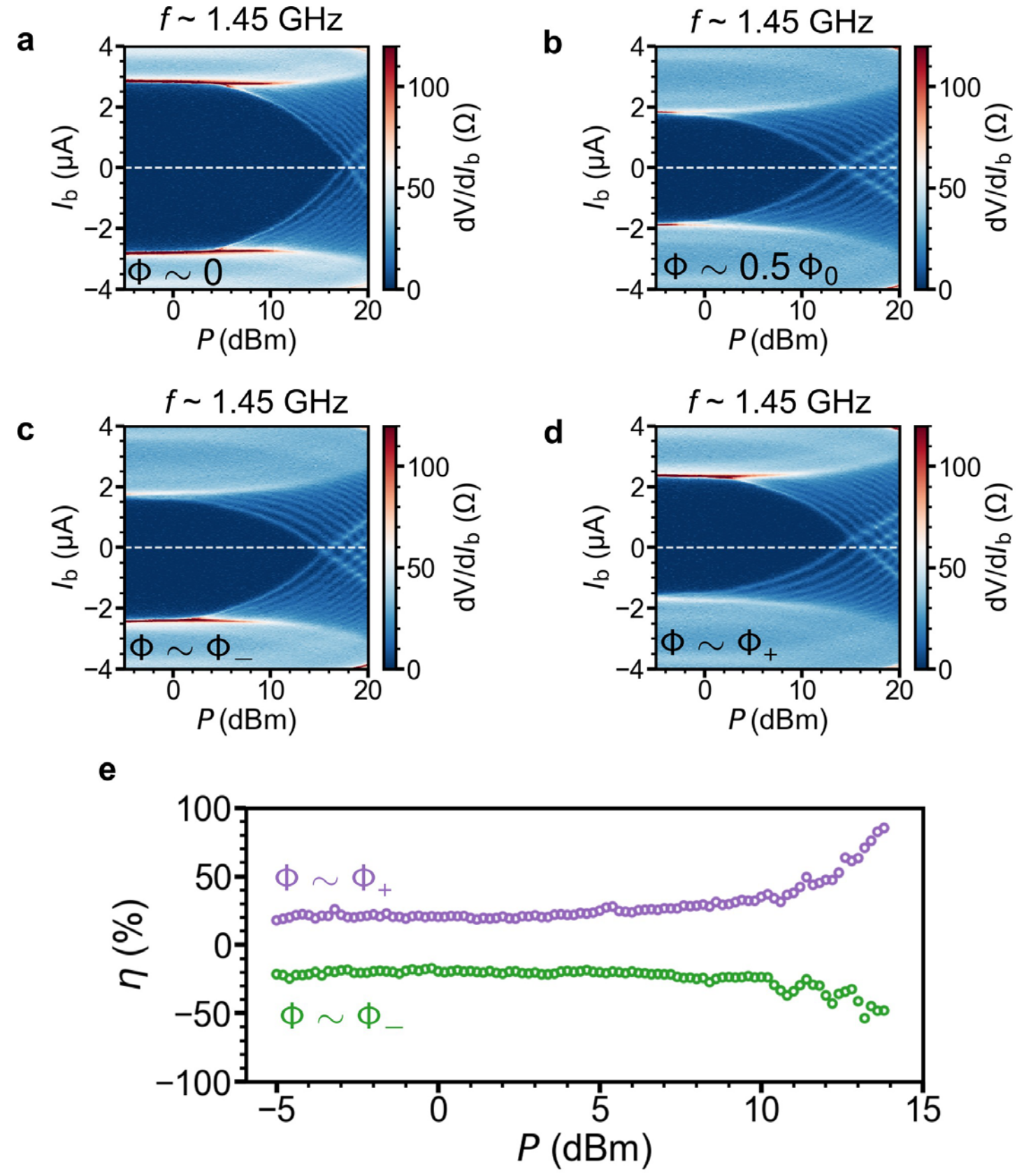

**Figure S7. Microwave response of the SDE in SQUID-2 under microwave irradiation with a frequency $f$ =1.45 GHz (adiabatic driving regime). a–d** Differential resistance $dV/dI_{\mathrm{b}}$ of SQUID-2 as a function of radiation power $P$ and bias current $I_{\mathrm{b}}$, measured at $\Phi\sim0$, $0.5\,\Phi_0$, $\Phi_-$ and $\Phi_+$, respectively. The microwave frequency is set at 1.45 GHz. **e** Diode efficiencies $\eta$ as a function of microwave power $P$. Green and purple circles correspond to data extracted from the measurements at $\Phi\sim\Phi_-$ and $\Phi\sim\Phi_+$, respectively.

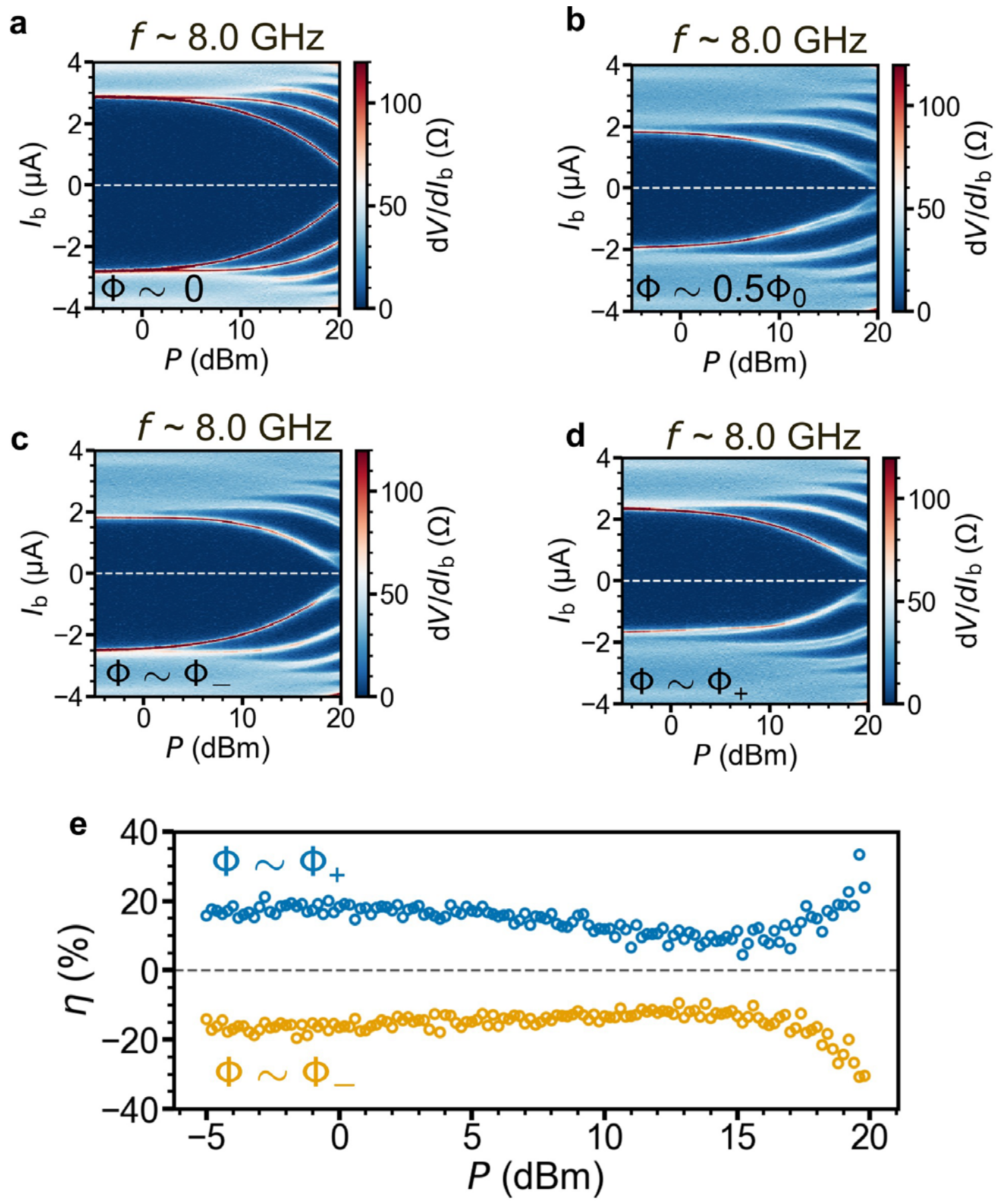

**Figure S8. Microwave response of the SDE in SQUID-2 under microwave irradiation with a frequency $f$ = 8.0 GHz (nonadiabatic driving regime). a–d** Differential resistance $dV/dI_{\mathrm{b}}$ of SQUID-2 as a function of irradiation power $P$ and bias current $I_{\mathrm{b}}$, measured at $\Phi \sim 0$, $0.5\,\Phi_0$, $\Phi_-$ and $\Phi_+$, respectively. The microwave frequency is $f \sim 8.0$ GHz. **e** Diode efficiencies $\eta$ as a function of $P$ extracted from **c** (yellow circles) and **d** (blue circles).

# Section 3. Theoretical Analysis

### 3.1 Fitting of microwave-driven switching currents in SQUID-1

The CPRs for JJ1 and JJ2 of a SQUID are given by:

$$I^j(\varphi) = \sum_{m=1} a_m^j \sin\left(m\varphi + \gamma_m^j\right) \tag{1}$$

Here, coefficients $a_{\mathrm{m}}^{\mathrm{j}}$ denotes the coefficient of the $m$th harmonics in the CPR of junction $j$, with $j = 1$ and 2 referring to JJ1 and JJ2, respectively. The phase shift of the $m$th harmonics of junction $j$ is denoted by $\gamma_{\mathrm{m}}^{\mathrm{j}}$. The total current $I(\varphi)$ passing through the SQUID is the sum of the currents from JJ1 and JJ2:

$$I = I^1(\varphi) + I^2\left(\varphi + 2\pi\frac{\Phi}{\Phi_0}\right) = \sum_{k\geq 1} I_{\mathrm{k}} \sin(k\varphi + \gamma_{\mathrm{k}}), \tag{2}$$

where $\Phi_0 = \frac{h}{2e}$ is the superconducting flux quantum. $I_{\mathrm{k}}$ and $\gamma_{\mathrm{k}}$ are the amplitude and phase shift of the $k$th harmonic of the total SQUID CPR, respectively. Following the procedure described in Refs. 1 and 2, we use the high-frequency expression for the time-averaged dc supercurrent $\bar{I}$ under an ac current drive of amplitude $I_{\mathrm{ac}}$:

$$\bar{I}(\alpha_0) = \sum_{k\geq 1} I_{\mathrm{k}} \sin(k\alpha_0 + \gamma_{\mathrm{k}}) J_0\left(k\frac{2eR_{\mathrm{n}}I_{\mathrm{ac}}}{\hbar\omega}\right). \tag{3}$$

Here, $J_0$ is the zeroth-order Bessel function, $\alpha_0$ is the time-independent component of the superconducting phase. The positive and negative critical currents correspond to the maximum and minimum dc currents as a function of $\alpha_0$. $R_{\mathrm{n}}$ is the normal-state resistance, and $\omega$ is the angular frequency of the ac drive.

Using this procedure, we simultaneously fit the positive and negative switching-current boundaries extracted from the four Shapiro maps of SQUID-1 measured at $f \sim 9.2$ GHz, as shown in Fig. 3 in the main article and Figure S2 in Supplementary Section 1. A single set of CPR parameters are used for the two junctions in all four maps. The fitting is restricted to the low-power region, $P <$ 6 dBm, where heating and possible nonequilibrium effects are expected to be relatively small. The extracted fitting parameters are listed in Table 1 and the corresponding fitting results are shown in Fig. S9. Only small variations ($\lesssim$ 10%) are found when varying the fitting range and/or including higher harmonics in the fitting procedure.

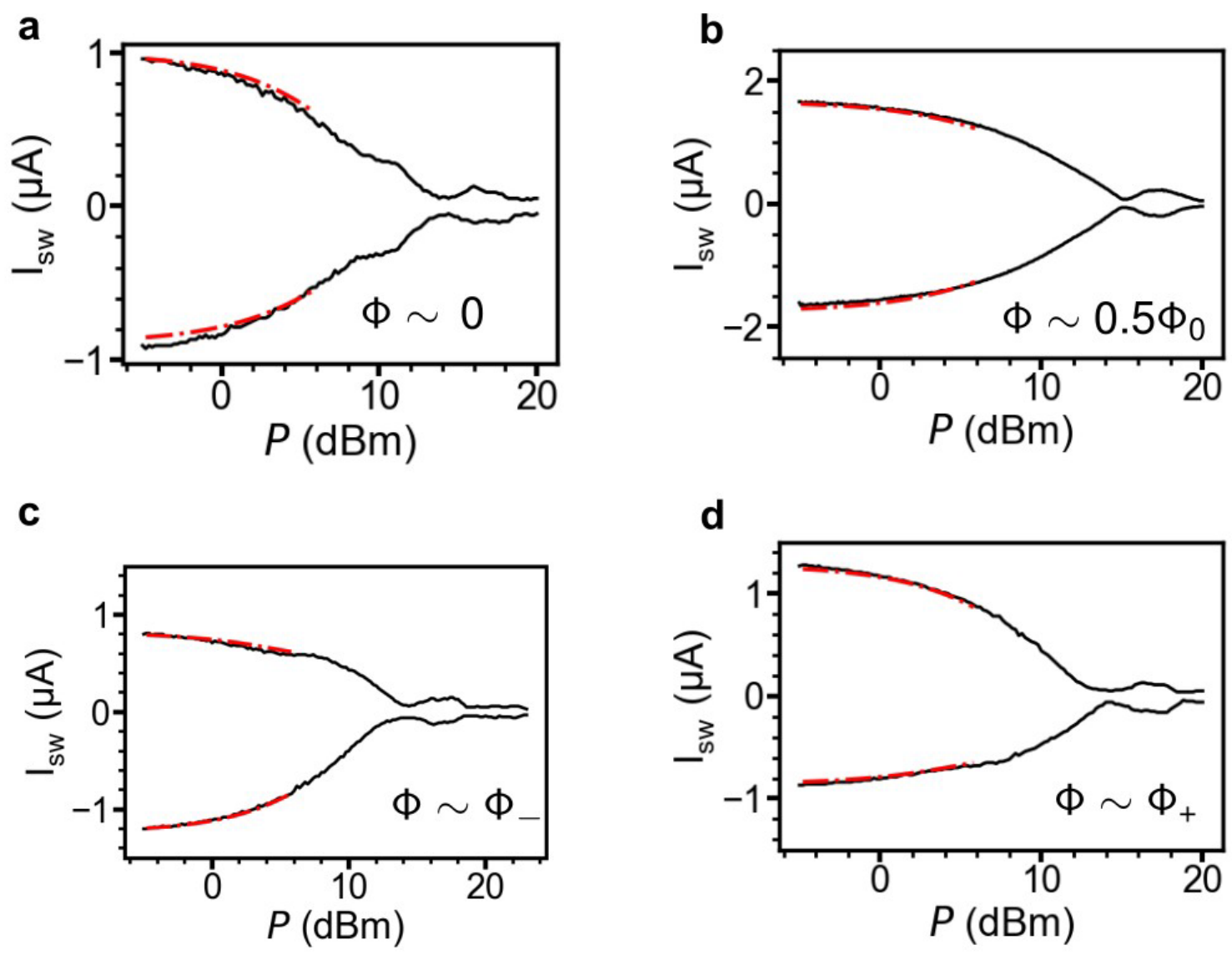

**Figure S9. Fitting of the switching currents in SQUID-1 under microwave irradiation. a-d** Black lines: Switching current $I_{\mathrm{sw}}$ as a function of irradiation power $P$ measured at $\Phi \sim 0$, $0.5\,\Phi_0$, $\Phi_-$ and $\Phi_+$, respectively. Data were extracted from Fig. 3 in the main article and Figure S2 in Supplementary Section 1. Red dash-dot lines: Fitting curves.

**Table 1: Fitting parameters for SQUID-1 under microwave irradiation**

| m | 1 | 2 |
|---|---|---|
| $a_m^1$(μA) | 1.043 | -0.435 |
| $a_m^2$(μA) | 0.452 | -0.057 |
| $\gamma_m^1$ | 0 | 0.509 |
| $\gamma_m^2$ | 0.210 | 0.447 |

### 3.2 Fitting of the magnetic-field/flux dependence of the switching currents in SQUID-1 without microwave irradiation

Alternatively, the CPRs of the two junctions can be extracted by fitting the switching current as a function of the perpendicular magnetic field measured without microwave irradiation. In this case, the fitting is highly sensitive to the initial parameter values because different CPRs can produce mathematically equivalent switching-current characteristics. Using the parameters listed in Table 1 as initial values, we fit the switching currents of SQUID-1 measured without microwave irradiation using the same data set as that shown in Fig. 1d in the main article. The resulting CPR parameters are listed in Table 2 and are in good agreement with those obtained from the Shapiro-map fits in Table 1. The corresponding fitting results are shown in Fig. S10.

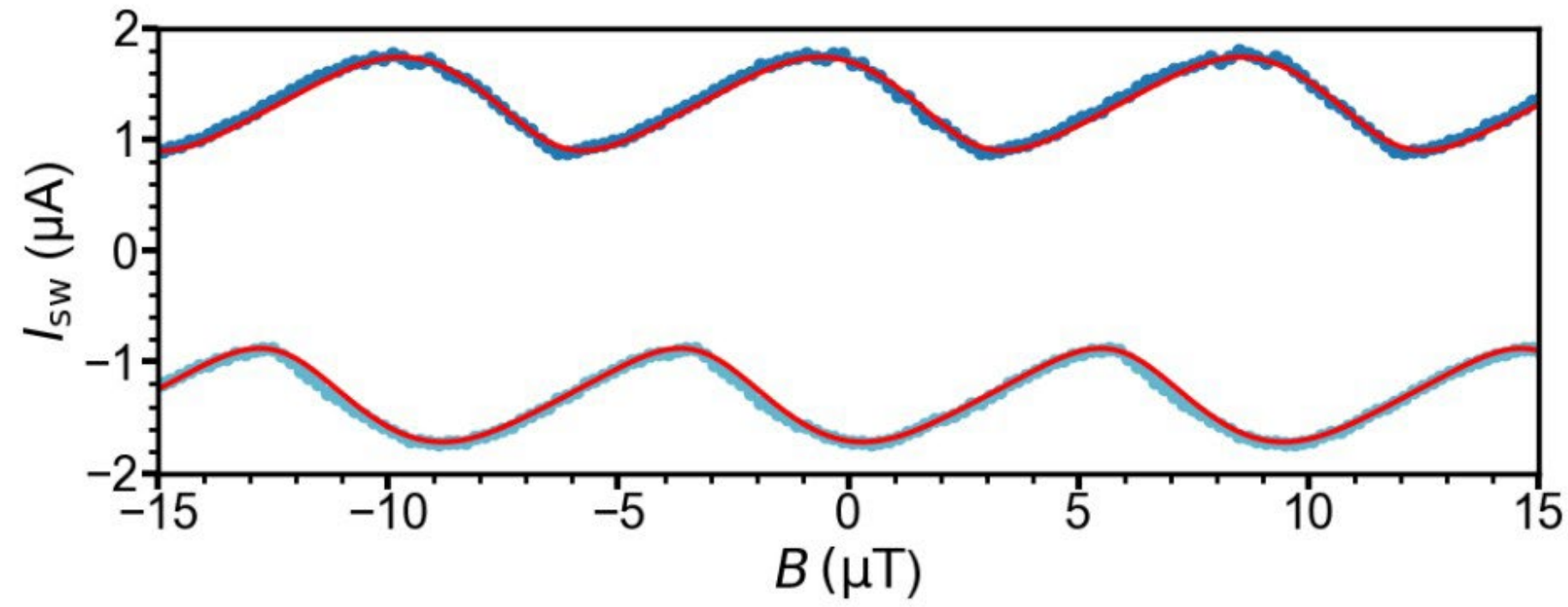

**Figure S10. Fitting of switching currents as a function of magnetic field for SQUID-1 based on the $I_b - \Phi$ characteristic measurements under no microwave irradiation.** Blue scatters: Switching current $I_{sw}$ as a function of external magnetic field. These data were extracted from the same measurement dataset as that used in plot for Fig. 1d in the main article. Red lines: Fitting curves.

**Table 2: Fitting parameters for SQUID-1 based on the $I_b - \Phi$ characteristic measurements under no microwave irradiation**

| m | 1 | 2 |
|---|---|---|
| $a_m^1$(μA) | 1.006 | -0.504 |
| $a_m^2$(μA) | 0.411 | -0.046 |
| $\gamma_m^1$ | 0 | 0.017 |
| $\gamma_m^2$ | 0.096 | 0.274 |

### 3.3 Simulated results obtained using the CPR parameters of SQUID-1 given in Table 1

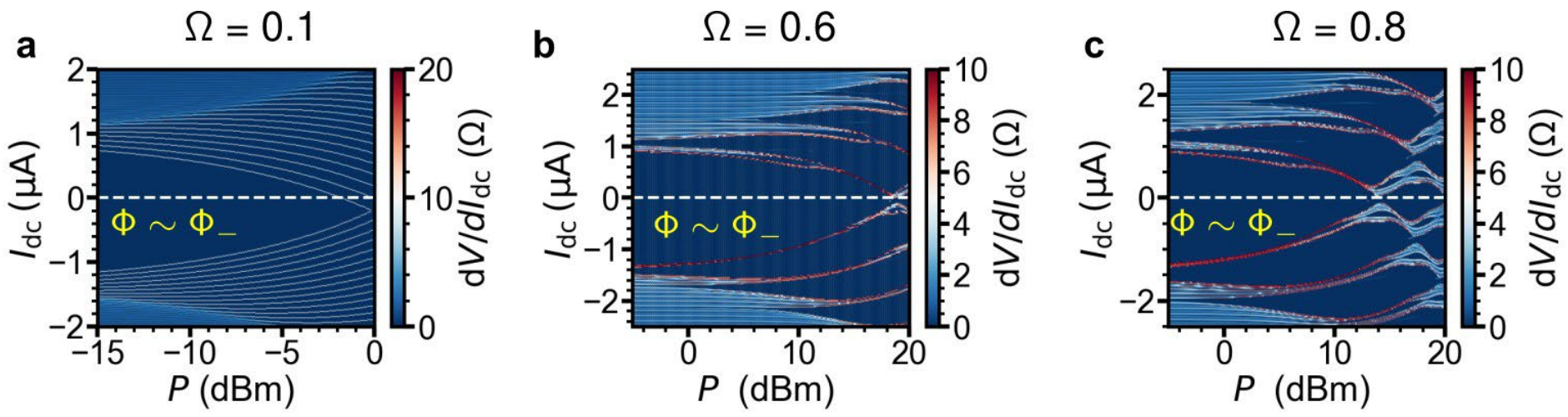

**Figure S11. Simulated Shapiro maps at $\Phi = \Phi_-$ for different microwave driving frequencies. a-f** Shapiro maps simulated using the extracted Josephson junction CPRs of SQUID-1 at $\Phi_-$. The reduced frequencies are $\Omega = 0.1, 0.6,$ and $0.8$ for panels **a**, **b**, and **c**, respectively.

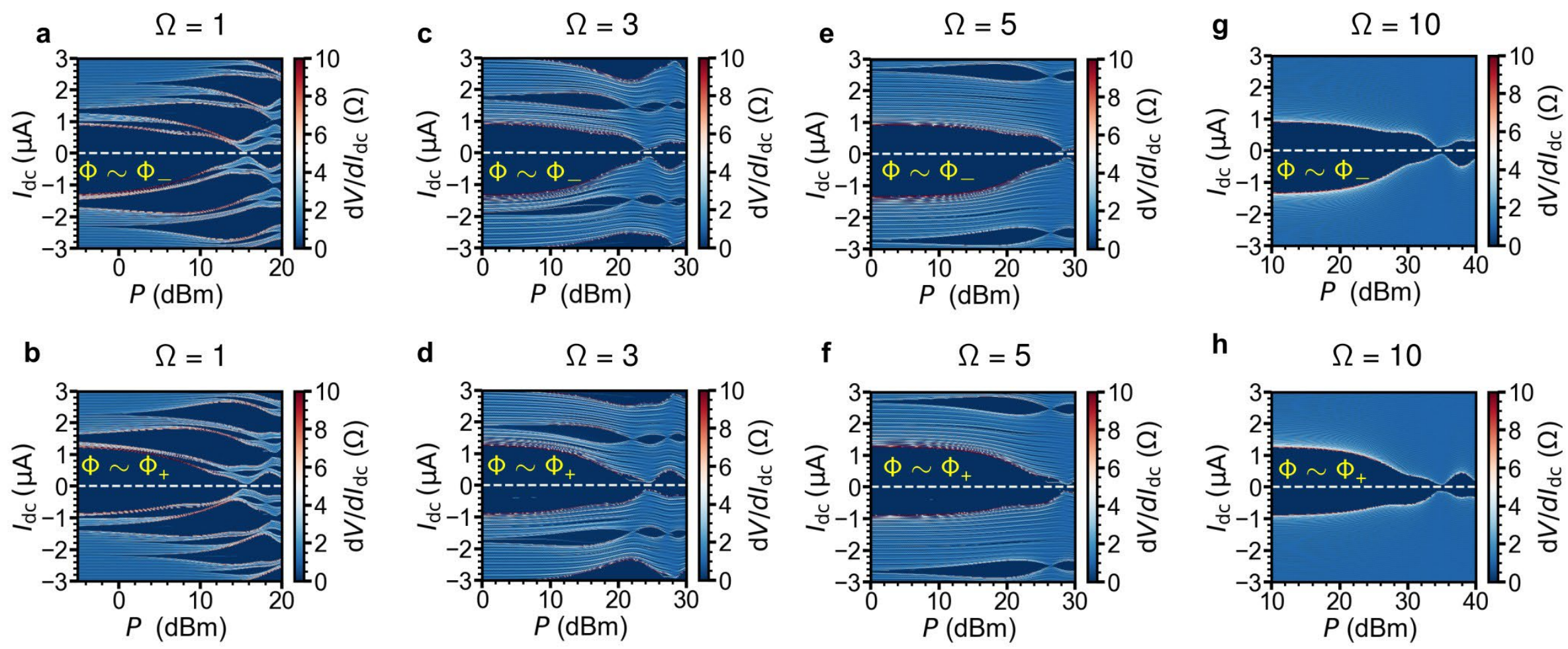

**Figure S12. Additional simulation results for SQUID-1 at frequencies higher than those accessible experimentally. a-h** Shapiro maps simulated using the extracted Josephson junction CPRs of SQUID-1 at $\Phi_-$ and $\Phi_+$. The reduced frequencies are $\Omega = 1$ (**a**, **b**), $3$ (**c**, **d**), $5$ (**e**, **f**) and $10$ (**g**, **h**).

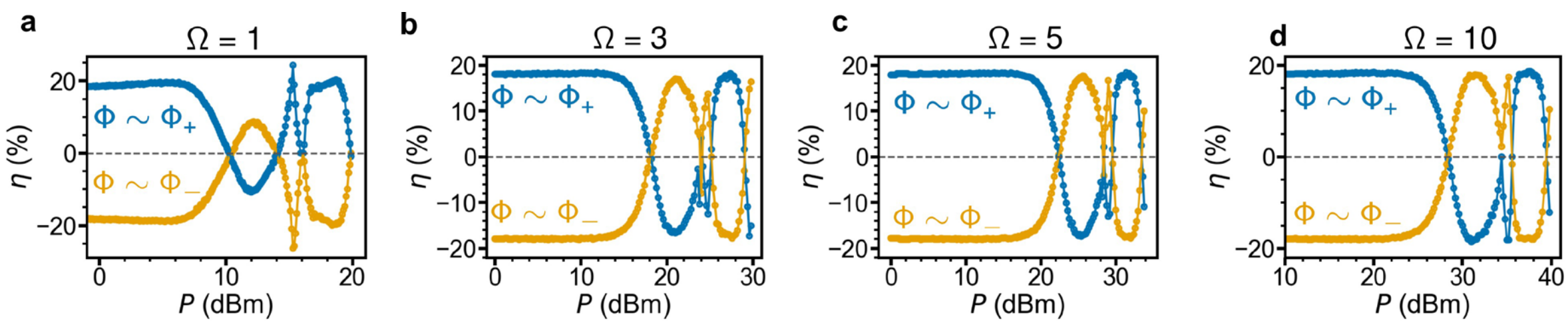

**Figure S13. a-d** Extracted diode efficiency $\eta(P)$ of SQUID-1 from the simulation results shown in Fig. S12. The reduced frequencies are $\Omega = 1, 3, 5,$ and $10$ for panels **a**, **b**, **c**, and **d**, respectively.

## Section 4. Estimation of Loop Inductance of SQUID-1

The loop inductance of SQUID-1 is estimated as,

$$L_{loop} = \frac{\mu_0 l}{2\pi}\left(ln\frac{2S}{rl} - 0.15 + \frac{\zeta}{4}\right),$$

where $\mu_0$ is the vacuum permeability, $S$ is the area enclosed by the SQUID loop, $l$ is the circumference of the loop, $r$ is the effective radius of the superconducting strip, and $\zeta$ is a parameter that takes the current distribution in the strip cross section into account. For SQUID-1, we obtain approximately $l \approx 81\ \mu\text{m}$, $S \approx 205\ \mu\text{m}^2$ and $r \approx 0.35\ \mu\text{m}$. Using $\zeta \approx 1$, we have $L_{loop} \approx 45$ pH.

The sheet kinetic inductance of the superconducting film is estimated as,

$$L_{\text{kin},\square} = \frac{hR_\square}{2\pi^2\Delta_0},$$

where $R_\square$ is the normal-state sheet resistance of the superconducting film and $\Delta_0$ is the superconducting gap. Using $R_\square \approx 1.9\ \Omega$ and assuming $\Delta_0 \approx 180\ \mu\text{eV}$, we obtain the kinetic inductance of SQUID-1 loop $L_{\text{kin}} \approx 256\ \text{pH}$. The total loop inductance is thus estimated to be $L_{\text{tot}} \approx 301$ pH.

## References

1. Seoane Souto, R.; Leijnse, M.; Schrade, C.; Valentini, M.; Katsaros, G.; Danon, J. Tuning the Josephson diode response with an ac current. *Phys. Rev. Res.* 2024, **6**, L022002.
2. Tjøtta, M.; Shah, D.; Modi, K.; Valentini, M.; Souto, R. S.; Katsaros, G.; Danon, J. Full Shapiro spectroscopy of current-phase relationships. *arXiv:*2511.14313 2025**.**